\documentclass[
    reprint, twocolumn,
    aps,prl,10pt,amsmath,amssymb,
    longbibliography,superscriptaddress,
    showpacs,preprintnumbers]{revtex4-1}

\usepackage{xr}
\makeatletter
\newcommand*{\addFileDependency}[1]{
  \typeout{(#1)}
  \@addtofilelist{#1}
  \IfFileExists{#1}{}{\typeout{No file #1.}}
}
\makeatother

\newcommand*{\myexternaldocument}[1]{
    \externaldocument{#1}
    \addFileDependency{#1.tex}
    \addFileDependency{#1.aux}
}

\myexternaldocument{supp}

\usepackage[colorlinks=true,citecolor=blue,linkcolor=blue]{hyperref}
\usepackage{oubraces}
\usepackage{bm,bbm,mathbbol}
\usepackage{physics}
\usepackage{mathtools}
\usepackage{amsmath, amsthm, amssymb, amsfonts}
\usepackage{bbold}
\usepackage{esvect}

\usepackage{epsfig}
\usepackage{array}
\usepackage{multirow}
\usepackage{graphicx}
\usepackage[normalem]{ulem}

\usepackage{float}
\usepackage[section]{placeins}
\usepackage{afterpage}

\usepackage{lipsum}

\usepackage{xifthen}

\makeatletter
\newcommand\footnoteref[1]{\protected@xdef\@thefnmark{\ref{#1}}\@footnotemark}
\makeatother

\newcommand{\nnnl}{\nonumber \\}

\newcommand{\myeqref}[1]{Eq.~\eqref{#1}}

\newcommand{\mb}[1]{\boldsymbol{\mathbf{#1}}}

\newcommand{\ansatz}[0]{\textit{ansatz}}
\newcommand{\Vtilde}[0]{{\widetilde{V}}}
\newcommand{\mbx}[0]{{\mb{x}}}
\newcommand{\oprho}[0]{{\hat{\rho}}}
\newcommand{\opH}[0]{{\hat{H}}}
\newcommand{\opV}[0]{{\hat{V}}}
\newcommand{\opU}[0]{{\hat{U}}}
\newcommand{\opUd}[0]{{\hat{U}^\dagger}}
\newcommand{\opr}[0]{{\hat{r}}}
\newcommand{\opp}[0]{{\hat{p}}}
\newcommand{\oprtilde}[0]{{\hat{\widetilde{r}}}}
\newcommand{\opptilde}[0]{{\hat{\widetilde{p}}}}
\newcommand{\opa}[0]{{\hat{a}}}
\newcommand{\opad}[0]{{\hat{a}^\dagger}}

\newcommand{\one}[0]{{\mathbb{1}}}
\newcommand{\hpmu}[0]{{\hphantom{\mu}}}

\newcommand{\dfdx}[2]{\frac{\partial {#1}}{\partial {#2}}}

\newcommand{\dfdxy}[3]{\frac{\partial^2 {#1}}{\partial {#2} \partial{#3}}}

\newcommand{\mbmcG}[1][]{
\ifthenelse{\isempty{#1}}{\mb{\mathcal{G}}}{\mb{\mathcal{G}}^{({#1})}}
}
\newcommand{\mbom}[0]{{\mb{\omega}}}
\newcommand{\mbPI}[0]{{\mb{P}_{1}}}
\newcommand{\mbPII}[0]{{\mb{P}_{2}}}
\newcommand{\mbHz}[0]{{\mb{H}^{(0)}}}
\newcommand{\mbVt}[0]{{\mb{V}^{(3)}}}
\newcommand{\mbVf}[0]{{\mb{V}^{(4)}}}
\newcommand{\mbPht}[0]{\mb{\Phi}^{(3)}}
\newcommand{\mbPhf}[0]{\mb{\Phi}^{(4)}}

\begin{document}

\title{Gaussian time-dependent variational principle for the finite-temperature anharmonic lattice dynamics}

\author{Jae-Mo Lihm}
\email{jaemo.lihm@gmail.com}
\author{Cheol-Hwan Park}
\email{cheolhwan@snu.ac.kr}
\affiliation{Center for Correlated Electron Systems, Institute for Basic Science, Seoul 08826, Korea}
\affiliation{Department of Physics and Astronomy, Seoul National University, Seoul 08826, Korea}
\affiliation{Center for Theoretical Physics, Seoul National University, Seoul 08826, Korea}

\date{\today}

\begin{abstract}
The anharmonic lattice is a representative example of an interacting, bosonic, many-body system.
The self-consistent harmonic approximation has proven versatile for the study of the equilibrium properties of anharmonic lattices.
However, the study of dynamical properties therein resorts to an {\ansatz}, whose validity has not yet been theoretically proven.
Here, we apply the time-dependent variational principle, a recently emerging useful tool for studying the dynamic properties of interacting many-body systems, to the anharmonic lattice Hamiltonian at finite temperature using the Gaussian states as the variational manifold.
We derive an analytic formula for the position-position correlation function and the phonon self-energy, proving the dynamical {\ansatz} of the self-consistent harmonic approximation.
We establish a fruitful connection between time-dependent variational principle and the anharmonic lattice Hamiltonian, providing insights in both fields.
Our work expands the range of applicability of the time-dependent variational principle to first-principles lattice Hamiltonians and lays the groundwork for the study of dynamical properties of the anharmonic lattice using a fully variational framework.
\end{abstract}

\footnotetext[1]{See Supplemental Material, which includes Refs.~\cite{PathriaBeale_StatMech},
at [URL will be inserted by publisher] for
the analysis of the variational parameters,
technical details of the derivations,
a note on degeneracies,
a note on the zero-temperature case,
and the calculation of the excitation energy of the single-mode anharmonic Hamiltonian.
}
\newcommand{\citeSupp}[0]{Note1}

\maketitle

\textit{Introduction} ---
Variational methods form the basis of our understanding of quantum mechanical many-body systems.
In a variational method, the wavefunctions or density matrices of a system are parametrized by a set of parameters whose size is much smaller than the dimension of the Hilbert space.
Static and time-dependent~\cite{1930DiracTDVP,2008KramerTDVP,2011HaegemanTDVP} variational methods are being actively used to study interacting many-body model Hamiltonians~\cite{2013Haegeman,2018AshidaVar,2018ShiNonGaussian,2019GuitaBosonic,2019RiveraVar,2019Vanderstraeten, 2019VanderstraetenReview,2020ShiTemp,2020WangVar,2020HacklReview}.

The anharmonic lattice is a representative example of an interacting bosonic many-body system in materials science.
The self-consistent harmonic approximation (SCHA) is a variational method for approximately finding the ground or thermal equilibrium state of an anharmonic lattice Hamiltonian~\cite{1955Hooton,2011Errea}.
Recently, a stochastic implementation of SCHA~\cite{2013Errea,2014Errea,2017Bianco,2018Monacelli} was developed and attracted considerable attention.
SCHA has been successfully applied to study
structural phase transitions~\cite{2017Bianco,2018Monacelli,2018BiancoPhase,2019AseginolazaPhase},
superconductivity~\cite{2013Errea,2015ErreaSC,2016ErreaSC,2017BorinagaSC,2020ErreaSC},
and charge density waves~\cite{2015LerouxCDW,2019BiancoCDW,2020ZhouCDW,2020BiancoCDW,2020SkyZhouCDW,2020DiegoCDW},
as well as to the dynamical properties such as the phonon spectral function~\cite{2015Paulatto,2018BiancoPhase,2019AseginolazaPhase,2019AseginolazaSpectral,2020AseginolazaSpectral}
and infrared and Raman spectra~\cite{2020MonacelliRaman}.

However, SCHA is limited in that one needs to resort to a specific {\ansatz} to study the dynamical properties.
It is known that the SCHA ansatz for the position-position Green function is correct in the static limit of zero frequency and the perturbative limit of weak anharmonicity~\cite{2017Bianco}.
However, the validity of the SCHA ansatz in the nonperturbative and dynamic regime~\cite{2018BiancoPhase,2019AseginolazaPhase,2019AseginolazaSpectral,2020MonacelliRaman}, where the dynamical theory is most necessary, has not been theoretically justified.

In this Letter, we solve this important problem by applying the time-dependent variational principle (TDVP) with Gaussian variational states~\cite{2012WeedbrookGaussian,2014AdessoGaussian,2019GuitaBosonic,2020HacklReview} to the anharmonic lattice Hamiltonian at finite temperature.
Gaussian TDVP expands the static variational states of SCHA to states with nonzero momenta.
We use the linearized time evolution to derive the position-position correlation function and prove the SCHA dynamical ansatz.
We illustrate that the Gaussian TDVP is successful in describing the dynamics because it includes the 2-phonon states as true dynamical excitations.
Our work connects the TDVP theory, whose application was mostly focused on model Hamiltonians for cold atoms, with anharmonic lattice dynamics and the SCHA method.
Such connection gives fruitful results on both sides.
In TDVP theory, the linearized time evolution and the projected Hamiltonian method~\cite{2013Haegeman, 2018ShiNonGaussian, 2019Vanderstraeten} are two different ways to compute the excitation spectrum, whose superiority over the other varies across systems~\cite{2013Haegeman,2019GuitaBosonic,2020HacklReview}.
We use the anharmonic lattice model to show that only the linearized time evolution gives correct excitation energies in the perturbative limit.
On the SCHA side, we illustrate ways to systematically expand the SCHA theory by leveraging recent developments of non-Gaussian TDVP~\cite{2018ShiNonGaussian,2020WangVar,2020ShiTemp}.

\textit{Self-consistent harmonic approximation} ---
We briefly review the key results of SCHA.
Within the adiabatic Born-Oppenheimer approximation, the anharmonic lattice Hamiltonian is
\begin{equation} \label{eq:scha_ham_def}
    \hat{H} = \sum_{a=1}^{N} \frac{\opptilde^2_{a}}{2 M_a} + \Vtilde(\oprtilde_{1}, ..., \oprtilde_{N}).
\end{equation}
Here, $a$ is the combined index for atoms and Cartesian directions, $N = N_{\rm atm} \times d$ with $N_{\rm atm}$ and $d$ the numbers of the atoms and the spatial dimensions, respectively, $M_a$ the atomic mass, $\oprtilde_a$ and $\opptilde_a$ the position and momentum operators, and $\Vtilde$ the Born-Oppenheimer potential energy.
We set $\hbar = 1$.

In SCHA, the true thermal equilibrium state of the anharmonic Hamiltonian is approximated by that of a harmonic Hamiltonian $\opH^{\rm (H)}$:
\begin{equation} \label{eq:scha_Hhar_def}
    \opH^{\rm (H)} = \sum_{a=1}^{N} \frac{\opptilde^2_{a}}{2 M_a} + \widetilde{V}^{\rm (H)}(\hat{\widetilde{\mathbf{r}}}).
\end{equation}
Since we study the dynamics around the SCHA equilibrium, we assume that the optimal harmonic potential $\widetilde{V}^{\rm (H)}$ is already found.
The SCHA density matrix is
\begin{equation} \label{eq:scha_rho0_def}
    \oprho_0 = e^{-\beta \opH^{\rm (H)}} / \Tr e^{-\beta \opH^{\rm (H)}},
\end{equation}
where $\beta = 1 / k_{\rm B} T$ is the inverse temperature.
For later use, we define $\expval{\hat{A}}_0 \equiv \Tr(\oprho_0 \hat{A})$.

Hereafter, we use the normal-mode representation, where the SCHA harmonic Hamiltonian becomes
\begin{equation} \label{eq:scha_Hhar_mode}
    \opH^{\rm (H)}
    = \sum_{m=1}^{N} \frac{\omega_m}{2} (\opp_{m}^2 + \opr_m^2),
\end{equation}
with $\omega_m$ the eigenvalue of the SCHA dynamical matrix, and $\opr_m$ and $\opp_m$ the normal-mode position and momentum operators.
The anharmonic Hamiltonian [Eq.~\eqref{eq:scha_ham_def}] can be written as
\begin{equation} \label{eq:scha_ham_mode_def}
    \hat{H} = \sum_{m=1}^{N} \frac{\omega_m}{2} \opp_{m}^2 + V(\mb{\opr}),
\end{equation}
with $V(\mb{\opr}) = \widetilde{V}(\mb{\oprtilde})$ the potential energy in the normal-mode representation.

In the normal-mode representation, the SCHA self-consistency equations~\cite{2017Bianco} become
\begin{equation} \label{eq:scha_dv}
    \expval{\dfdx{\opV}{r_m}}_0 = 0,\quad  \expval{\dfdxy{\opV}{r_m}{r_n}}_0 = \omega_m \delta_{m,n}.
\end{equation}
Also, since $\oprho_0$ is a thermal state, we find
\begin{equation} \label{eq:scha_p}
    \expval{\opp_m}_0 = 0,\quad \expval{\opp_m \opp_n}_0 = \left( n_m + \frac{1}{2} \right) \delta_{m,n},
\end{equation}
with $n_m = 1 / (e^{\beta \omega_m} - 1)$ the occupation number.

\textit{Gaussian time-dependent variational principle} ---
Next, we discuss the general principles of Gaussian TDVP for a multimode bosonic system at finite temperature.
We use the set of states obtained by applying a Gaussian unitary transformation $\opU(\mbx)$ to the SCHA density matrix as the variational manifold:
\begin{equation}
    \oprho(\mbx) = \opU(\mbx) \oprho_0 \opUd(\mbx).
\end{equation}
Here, $\mbx$ is a real-valued vector that encodes all the variational parameters.
We parametrize the Gaussian transformation as
\begin{equation} \label{eq:tdvp_U_def}
    \hat{U}(\mbx) = \hat{D}(\mb{\alpha}) \hat{S}(\mb{\beta}, \mb{\gamma}),
\end{equation}
where $\hat{D}$ and $\hat{S}$ are the displacement and squeezing operators, respectively:
\begin{align} \label{eq:tdvp_D_def}
\hat{D}(\mb{\alpha})
=& \exp( \frac{1}{\sqrt{2}} \sum_m (\alpha_m \opad_m - \alpha_m^* \opa_m) ),
\end{align}
\begin{align} \label{eq:tdvp_S_def}
\hat{S}(\mb{\beta},\mb{\gamma})
= {\rm exp} \Bigg[
& \sum_{\substack{m,n\\m \leq n}} b_{mn}
(\beta_{mn}\opad_m \opad_n - \beta_{mn}^* \opa_m \opa_n ) \nnnl
+& \sum_{\substack{m,n\\m < n}} c_{mn} (
\gamma_{mn}^*\opad_m \opa_n - \gamma_{mn} \opad_n \opa_m) \Bigg],
\end{align}
where
\begin{equation} \label{eq:s_tdvp_bmn_def}
b_{mn} \equiv \begin{cases} 1 / \sqrt{4(n_m + n_n + 1)} & \text{if } m = n \\ 1 / \sqrt{2(n_m + n_n + 1)} & \text{if } m \neq n \end{cases},
\end{equation}
\begin{equation} \label{eq:s_tdvp_cmn_def}
c_{mn} \equiv 1 / \sqrt{2(n_m - n_n)}.
\end{equation}
The variational parameters $\alpha_m$, $\beta_{mn}$, and $\gamma_{mn}$ are complex numbers.
The parameter $\beta_{mn}$ ($\gamma_{mn}$) is defined only for $m \leq n$ ($m < n$).
Here, we assume for simplicity that $\omega_m$'s are nondegenerate and satisfy $\omega_1 < \omega_2 < \cdots < \omega_{N}$.
The total number of complex variational parameters is $N^2 + N$.
In the linear response regime, degeneracy does not pose any theoretical difficulty: if modes $m$ and $n$ are degenerate, one just needs to exclude $\gamma_{mn}$ from the set of variational parameters.
This exclusion is done because the infinitesimal transformation parametrized by $\gamma_{mn}$ does not change $\oprho_0$~\cite{\citeSupp}.

Each group of parameters describes a different type of excitation.
Parameters $\mb{\alpha}$, $\mb{\beta}$, and $\mb{\gamma}$ correspond to 1-phonon excitations, 2-phonon excitations with two creations or two annihilations of phonons, and 2-phonon excitations with one creation and one annihilation, respectively.

The imaginary parts of the parameters generate dynamics.
For example, $\Im \mb{\alpha}$ generates a finite atomic momentum through the displacement operator.
The SCHA theory does not contain these imaginary parameters because the variational states are limited to the thermal state of a harmonic Hamiltonian.
In contrast, Gaussian TDVP, which allows both the real and imaginary parts of the variational parameters to vary, naturally allows one to study the dynamics.

We define $\mbx$, the vector of variational parameters as
\begin{equation} \label{eq:tdvp_x_def}
    \mbx = (
    \Re \mb{\alpha} \ \Im \mb{\alpha} \ 
    \Re \mb{\beta} \ \Im \mb{\beta} \ 
    \Re \mb{\gamma} \ \Im \mb{\gamma} )^\intercal.
\end{equation}
Since $\oprho(\mbx = 0) = \oprho_0$ is the variational solution that minimizes the SCHA free energy, $\mbx = 0$ is a stationary point of the variational time evolution~\cite{\citeSupp}.

To apply TDVP to mixed states, we map the variational density matrices to wavefunctions by purification~\cite{2000Nielsen,2020ShiTemp}.
For each physical state in the number basis, we add an auxiliary state so that the purified wavefunction becomes
\begin{equation} \label{eq:pur_psi_var}
    \ket{\Psi(\mbx)} = [\opU(\mbx) \sqrt{\oprho_0} \otimes \one] \ket{\Phi^+},
\end{equation}
where $\otimes$ denotes a tensor product, and $\ket{\Phi^+}$ is a maximally entangled state~\cite{2000Nielsen} between the physical and the auxiliary modes [see \myeqref{eq:s_tdvp_Phi_def} and related discussions].
For the purified wavefunction, the expectation value of a physical operator $\hat{A}_0$ is
\begin{align} \label{eq:pur_e_expval}
    A(\mb{x})
    \equiv \mel{\Psi(\mb{x})}{\hat{A}_0 \otimes \one}{\Psi(\mb{x})}
    = \expval{\hat{A}(\mb{x})}_0,
\end{align}
where
\begin{equation} \label{eq:s_transf_op_x}
    \hat{A}(\mbx) = \opUd(\mbx) \hat{A}_0 \opU(\mbx).
\end{equation}

The variational time evolution is obtained by projecting the true dynamics of the wavefunction to the tangent space of the variational manifold.
The tangent space is spanned by the tangent vectors, which at $\mbx = 0$ are
\begin{align} \label{eq:tdvp_tan_def_2}
    \ket{V_\mu}
    =& \left[ (\partial_\mu \opU) \sqrt{\oprho_0} \otimes \one \right] \ket{\Phi^+}.
\end{align}

Using the variational linear response theory~\cite{2020HacklReview,\citeSupp}, one can show that the retarded correlation function $G_{AB}^{\rm (R)}(\omega)$ between operators $\hat{A}$ and $\hat{B}$ is
\begin{equation} \label{eq:tdvp_lr_eqn}
    G_{AB}^{\rm (R)}(\omega)
    = \lim_{\eta \rightarrow 0^+} -i (\partial_\mu B)  \mathcal{G}^\mu_{\hpmu\nu}(\omega + i\eta) (\Omega^{\nu\rho} \partial_\rho A).
\end{equation}
Here, the matrix Green function $\mbmcG(z)$ is defined as
\begin{equation} \label{eq:tdvp_G_def}
(z - i\mb{K}) \mbmcG(z) = \one,
\end{equation}
where $\mb{K}$ is the linearized time-evolution generator defined as
\begin{equation} \label{eq:tdvp_K_def}
K^\mu_{\hpmu\nu}
= -\Omega^{\mu\rho} \partial_\rho \partial_\nu E,
\end{equation}
with $E(\mbx) = \Tr[ \oprho(\mb{x}) \opH ]$.
The symplectic form $\mb{\Omega}$ is defined by
\begin{equation} \label{eq:tdvp_Omega_def}
    \Omega^{\mu \rho} \Im \braket{V_\rho}{V_\nu} = \frac{1}{2} \delta^{\mu}_{\hpmu\nu}.
\end{equation}
By computing $\mb{K}$ and the corresponding matrix Green function $\mbmcG(z)$, one can find the physical correlation function $G_{AB}^{\rm (R)}(\omega)$ using \myeqref{eq:tdvp_lr_eqn}.

\textit{Anharmonic lattice dynamics} ---
Now, we study the dynamical properties of the anharmonic lattice Hamiltonian using Gaussian TDVP.
First, the symplectic form is~\cite{\citeSupp}
\begin{equation} \label{eq:anh_Omega}
    \mb{\Omega} =
    \begin{pmatrix} 0 & -\one \\ \one & 0 \end{pmatrix}\oplus
    \begin{pmatrix} 0 & -\one \\ \one & 0 \end{pmatrix} \oplus
    \begin{pmatrix} 0 & -\one \\ \one & 0 \end{pmatrix},
\end{equation}
with $\oplus$ the direct sum.

The three matrices correspond to the subspace spanned by the tangent vectors for the variation of $\mb{\alpha}$, $\mb{\beta}$, and $\mb{\gamma}$, respectively.
In each matrix, the bases for the first (second) block of rows and columns are the tangent vectors for the real (imaginary) parts of the parameters.

For later use, we define $\mbPI$, $\mathbf{P}_{2+}$, and $\mathbf{P}_{2-}$ as the projection operators to the bases of each of the three matrices.
The subscripts $1$, $2+$, and $2-$ indicate the nature of the tangent vectors: 1-phonon excitations, 2-phonon excitations with two creations or two annihilations, and 2-phonon excitations with one creation and one annihilation.
We also define the projection to the whole 2-phonon sector: $\mbPII = \mathbf{P}_{2+} + \mathbf{P}_{2-}$.

Evaluating \myeqref{eq:tdvp_K_def}, we find that the time evolution generator $\mb{K}$ is the sum of the non-interacting part, 3-phonon interaction, and 4-phonon interaction (see Sec.~\ref{sec:supp_der_2} of the Supplementary Material~\cite{\citeSupp}):
\begin{equation}
i \mb{K} = \mb{H}^{(0)} + \mb{V}^{(3)} + \mb{V}^{(4)},
\end{equation}
where
\begin{equation} \label{eq:anh_H0_def}
\mb{H}^{(0)} =
\begin{pmatrix} 0 & i\mb{\omega} \\ -i\mb{\omega} & 0 \end{pmatrix} \oplus
\begin{pmatrix} 0 & i\mbom_{+} \\ -i\mbom_{+} & 0 \end{pmatrix} \oplus
\begin{pmatrix} 0 & i\mbom_{-} \\ -i\mbom_{-} & 0 \end{pmatrix},
\end{equation}
\begin{equation} \label{eq:anh_V3_def}
\mb{V}^{(3)}
= \begin{pmatrix}
0 & 0 & 0 & 0 & 0 & 0\\
0 & 0 & -i \mbPht B & 0 & -i \mbPht C & 0\\
0 & 0 & 0 & 0 & 0 & 0\\
-iB \mbPht & 0 & 0 & 0 & 0 & 0 \\
0 & 0 & 0 & 0 & 0 & 0\\
-iC \mbPht & 0 & 0 & 0 & 0 & 0
\end{pmatrix},
\end{equation}
\begin{equation} \label{eq:anh_V4_def}
\mb{V}^{(4)}
= \begin{pmatrix} 0 & 0 \\ 0 & 0 \end{pmatrix} \oplus
\begin{pmatrix}
0 & 0 & 0 & 0\\
-iB \mbPhf B& 0 & -i B \mbPhf C & 0 \\
0 & 0 & 0 & 0\\
-iC \mbPhf B & 0 & -iC \mbPhf C & 0
\end{pmatrix}.
\end{equation}
Here, we defined the diagonal matrices:
\begin{equation} \label{eq:anh_omega_def}
    \mb{\omega}_{m,n} = \omega_m \delta_{m,n},
\end{equation}
\begin{equation} \label{eq:anh_omegapp_def}
    [\mbom_{\pm}]_{mn,pq} = (\omega_m \pm \omega_n) \delta_{mn,pq},
\end{equation}
\begin{align} \label{eq:anh_B_def}
    B_{mn,pq}
    =& b_{mn} (n_m + n_n + 1) \delta_{mn,pq},
\end{align}
\begin{align} \label{eq:anh_C_def}
    C_{mn,pq}
    =& - c_{mn} (n_m - n_n) \delta_{mn,pq}.
\end{align}
The implicit summation over a pair of mode indices $m$ and $n$ implies the constraint $m \leq n$ unless otherwise noted.
We also defined the anharmonicity tensor
\begin{equation} \label{eq:anh_Phi_def}
    \Phi^{(m)}_{n_1, \cdots, n_m} = \expval{ \frac{\partial^m V}{\partial r_{n_1} \cdots \partial r_{n_m}} }_0.
\end{equation}
The non-interacting part $\mb{H}^{(0)}$ describes the free evolution of 1- and 2-phonon excitations in the SCHA Hamiltonian.
The 3-phonon interaction $\mb{V}^{(3)}$ couples the 1- and 2-phonon excitations.
The 4-phonon interaction $\mb{V}^{(4)}$ couples the 2-phonon excitations to each other.

Finally, we study the linear response of the anharmonic lattice and compute the position-position correlation function.
First, we define the non-interacting Green function $\mbmcG[0]$:
\begin{equation} \label{eq:anh_G0_green}
    (z - \mbHz) \mbmcG[0](z) = \one.
\end{equation}
From \myeqref{eq:anh_H0_def}, one finds
\begin{align} \label{eq:anh_G0}
    \mbmcG[0](z)
    = \mbmcG[0]_{1}(z) \oplus \mbmcG[0]_{2+}(z) \oplus \mbmcG[0]_{2-}(z),
\end{align}
where
\begin{equation} \label{eq:anh_G01_def}
    \mbmcG[0]_{1}(z)
    = \frac{1}{z^2 - \mbom^2}
    \begin{pmatrix}
    z & i\mbom \\
    -i\mbom & z
    \end{pmatrix},
\end{equation}
and
\begin{equation} \label{eq:anh_G02_def}
    \mbmcG[0]_{2\pm}(z)
    = \frac{1}{z^2 - \mbom_{\pm}^2}
    \begin{pmatrix}
    z & i\mbom_{\pm} \\
    -i\mbom_{\pm} & z
    \end{pmatrix}.
\end{equation}

Next, we include the 4-phonon interaction $\mb{V}^{(4)}$.
We define the partially interacting Green function $\mbmcG[4](z)$:
\begin{equation} \label{eq:anh_G4_def}
(z - \mb{H}^{(0)} - \mb{V}^{(4)}) \mbmcG[4](z) = \one.
\end{equation}
Since the 4-phonon interaction $\mb{V}^{(4)}$ does not act on the 1-phonon sector, we find
\begin{equation} \label{eq:anh_G4_1}
    \mbPI \mbmcG[4] \mbPI
    = \mbmcG[0]_{1} \oplus 0.
\end{equation}
For the 2-phonon sector, we obtain the Dyson equation
\begin{equation} \label{eq:anh_G4_2_dyson}
    \mbPII \mbmcG[4] \mbPII
    = \mbPII \mbmcG[0] \mbPII + \mbPII \mbmcG[4] \mbVf \mbmcG[0] \mbPII.
\end{equation}

Finally, we study the fully interacting Green function $\mbmcG(z)$ by including the 3-phonon interaction $\mbVt$.
From the definitions of $\mbmcG$ and $\mbmcG[4]$, we obtain the Dyson equation
\begin{align} \label{eq:anh_G1_Dyson_before}
&\mbPI \mbmcG \mbPI
= \mbPI \mbmcG[4] \mbPI \\
&+ \mbPI \mbmcG[4] \mbPI \mbVt \mbPII \mbmcG[4] \mbPII \mbVt \mbPI \mbmcG \mbPI. \nonumber
\end{align}
One can solve the Dyson equations [Eqs.~(\ref{eq:anh_G4_2_dyson}, \ref{eq:anh_G1_Dyson_before})] to find~\cite{\citeSupp}
\begin{align} \label{eq:anh_G1_Dyson}
&\mbPI \mbmcG \mbPI
= \mbmcG[0]_{1} \\
&- \mbmcG[0]_{1}
\begin{pmatrix} 0 & 0 \\ \sum\limits_{s,s'=\pm} \mbPht B_s [\mbmcG[4]_{ss'}]_{12} B_{s'} \mbPht & 0 \end{pmatrix}
\mbPI \mbmcG \mbPI. \nonumber
\end{align}
Here, we defined $B_+ = B$ and $B_- = C$.
In \myeqref{eq:anh_G1_Dyson}, we omitted the direct sum of the zero matrix in the $\mbPII$ subspace for brevity.

From Eqs.~(\ref{eq:s_transf_op_der}, \ref{eq:s_overlap_dU_ar}), one finds that the matrix elements for the position operator is nonzero only for the variation of $\Re \mb{\alpha}$:
\begin{equation}
    \partial_\mu \mb{r} = \begin{pmatrix} \one & 0 & 0 & 0 & 0 & 0 \end{pmatrix}^\intercal.
\end{equation}
Then, from Eqs.~(\ref{eq:tdvp_lr_eqn}, \ref{eq:anh_G1_Dyson}), one can derive the Dyson equation for the interacting retarded position-position correlation function~\cite{\citeSupp}:
\begin{equation} \label{eq:anh_Grr_Dyson}
    \mb{G}_{rr}^{\rm (R)} = \mb{G}_{rr}^{\rm (R0)}
    + \mb{G}_{rr}^{\rm (R0)} \mb{\Pi}_{rr} \mb{G}_{rr}^{\rm (R)}.
\end{equation}
The self-energy is
\begin{align} \label{eq:anh_Pi}
\mb{\Pi}_{rr}(z)
=& \mbPht \mb{W} (\one - \mbPhf \mb{W})^{-1} \mbPht
\end{align}
where $\mb{W}$ is a diagonal matrix defined as
\begin{align} \label{eq:anh_W_def}
    \mb{W} = \sum_{s=\pm} B_s \frac{\mb{\omega}_s}{z^2 - \mb{\omega}_s^2} B_s.
\end{align}

By recovering the mode indices and defining
\begin{align} \label{eq:anh_chi_def}
    \chi_{mn,pq}(z)
    \equiv \frac{1}{2} &\Big[ \frac{(\omega_m + \omega_n)(n_m + n_n + 1)}{(\omega_m + \omega_n)^2 - z^2} \nnnl
    &- \frac{(\omega_m - \omega_n)(n_m - n_n)}{(\omega_m - \omega_n)^2 - z^2} \Big] \delta_{mn,pq},
\end{align}
one can rewrite \myeqref{eq:anh_Pi} in a form identical to the SCHA dynamical ansatz~\cite{\citeSupp}:
\begin{equation} \label{eq:anh_Pi_chi}
    \mb{\Pi}_{rr}(z)
    = \mbPht
    \left( -\frac{1}{2} \mb{\chi}(z) \right)
    \left[ \one - \mbPhf \left( -\frac{1}{2} \mb{\chi}(z) \right) \right]^{-1}
    \mbPht.
\end{equation}
In \myeqref{eq:anh_Pi_chi}, the implicit summation over the mode indices is done without any constraints.
Equation \eqref{eq:anh_Pi_chi} and its derivation is the main result of this Letter.
When transformed to the Cartesian representation, \myeqref{eq:anh_Pi_chi} becomes identical to the SCHA dynamical ansatz [Eq.~(70) of Ref.~\citealp{2017Bianco}].
We emphasize that we rigorously derived the phonon self-energy $\mb{\Pi}_{rr}(z)$ using Gaussian TDVP.
Our derivation theoretically \textit{proves} the SCHA dynamical ansatz.

The physical interpretation of the self-energy formula we obtained vary significantly from that of the SCHA dynamical ansatz.
In Gaussian TDVP, the 2-phonon states are true dynamical excitations.
However, in SCHA, the 2-phonon states do not have their own dynamics and appear only indirectly through the position dependence of the SCHA force constants.
The presence of the dynamical 2-phonon excitations is the essential reason why Gaussian TDVP can describe dynamical properties while the SCHA theory cannot.

For example, the phonon lifetime is an important dynamical property of an anharmonic lattice.
In Gaussian TDVP, the 1-phonon states acquire a finite lifetime by decaying to the continuum of 2-phonon states through the 3-phonon interaction.
In contrast, in SCHA, there are no continuum states to which the 1-phonon states can decay.
Hence, in SCHA, the phonon lifetimes can only be described with a perturbative approximation~\cite{2015Paulatto} unless one resorts to an ansatz.

\textit{Discussion} ---
A common alternative to the linearized time evolution is the projected Hamiltonian method~\cite{2013Haegeman, 2018ShiNonGaussian, 2019Vanderstraeten}.
There, the Hamiltonian is projected onto the tangent space of the variational manifold.
Let us consider a single-mode anharmonic oscillator at $T=0$, whose Hamiltonian is
\begin{equation} \label{eq:single_mode_ham}
    \opH = \frac{\omega_0}{2} (\opp^2 + \opr^2)
    + \frac{\lambda a}{6} \left( \opr^3 - \frac{3}{2} \opr \right)
    + \frac{\lambda^2 b}{24} \left( \opr^4 - 3\opr^2 + \frac{3}{4} \right).
\end{equation}
Here, $\lambda$ is the perturbation strength.
The SCHA variational Hamiltonian is
\begin{equation} \label{eq:single_mode_scha}
    \opH^{\rm (H)} = \frac{\omega_0}{2} (\opp^2 + \opr^2),
\end{equation}
and the variational ground-state energy is $\omega_0 / 2$.

\begin{table}
\begin{tabular}{ c | c }
 \hline
 Perturbation theory &
 $\omega_0 - \lambda^2 a^2 / 12 \omega_0 + \mathcal{O}(\lambda^4)$ \\
 Linearized time evolution &
 $\omega_0 - \lambda^2 a^2 / 12 \omega_0 + \mathcal{O}(\lambda^4)$ \\
 Projected Hamiltonian &
 $\omega_0 - \lambda^2 a^2 / 16 \omega_0 + \mathcal{O}(\lambda^4)$ \\
 \hline
\end{tabular}
\caption{\label{table:excitation} Excitation energy of the anharmonic oscillator~[\myeqref{eq:single_mode_ham}] computed with three different methods.}
\end{table}

In Table~\ref{table:excitation} we list the excitation energy, the difference of the ground- and first-excited state energy, computed using different methods~\cite{\citeSupp}.
Comparing the variational methods to the perturbation theory, we find that the linearized time evolution is correct in the perturbative limit $\lambda \rightarrow 0$, while the projected Hamiltonian method is not.
Since the SCHA dynamical ansatz is exact in the perturbative limit~\cite{2017Bianco}, this finding also holds for a general multimode anharmonic lattice at finite temperatures.

\begin{figure}
\includegraphics[width=1.0\columnwidth]{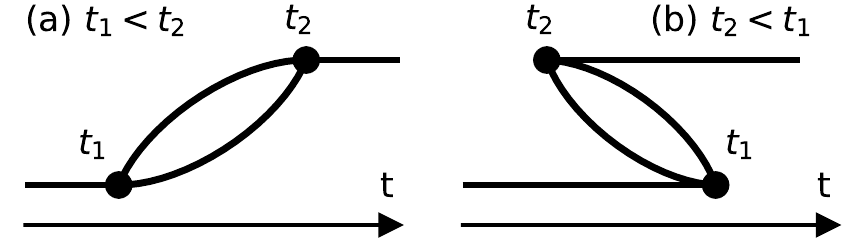}
\caption{
Diagrams of the two processes that appear in the time domain representation of a bubble diagram.
Created using the \texttt{feynman} package~\cite{AntoniusFeynman}.
}
\label{fig:bubble}
\end{figure}

This difference occurs because the projected Hamiltonian method fails to describe the effect of virtual 3- and 4-phonon states.
In Fig.~\ref{fig:bubble}, we show the two processes that appear in the time domain representation of the bubble diagram for the phonon self-energy.
Figure~\ref{fig:bubble}(b) describes a process involving a 4-phonon state.
Since the Gaussian projected Hamiltonian method completely neglects the 3- and 4- phonon excitations, it only includes the process described in Fig.~\ref{fig:bubble}(a), not that of Fig.~\ref{fig:bubble}(b).
In contrast, in the linearized time evolution, the coupling of the 1- and 2-phonon states to virtual 3- and 4-phonon states is included by an additional term related to the derivative of the tangent vectors, which is neglected in the projected Hamiltonian method~\cite{2020HacklReview}.
Thanks to this additional term, the linearized time evolution gives the correct perturbative limit, while the projected Hamiltonian cannot.

A promising future research direction based on our study is a rigorous, systematic expansion of the SCHA method to go beyond the harmonic approximation by using non-Gaussian variational transformations~\cite{2018ShiNonGaussian}.
Also, the use of mixed fermionic and bosonic variational states~\cite{2018ShiNonGaussian,2020WangVar,2020ShiTemp}
will allow the study of nontrivial electron-phonon correlation such as in phonon-mediated superconductivity or polarons in anharmonic lattices.

Recently, Monacelli and Mauri also reported a proof of the SCHA dynamical self-energy in an independent work~\cite{2020MonacelliTDSCHA}.
While Ref.~\cite{2020MonacelliTDSCHA} additionally presents a numerical algorithm to compute the correlation functions,
our work focuses on the link between TDVP and SCHA.
Also, while the proof for the finite-temperature case in Ref.~\cite{2020MonacelliTDSCHA} is based on an analogy with the $T$=0 case, our proof uses purification to rigorously derive the finite-temperature equation of motion.
The results of the two works are consistent when there is an overlap.

\textit{Conclusion} ---
In summary, we developed a variational theory for the dynamical properties of anharmonic lattices using Gaussian TDVP, establishing a firm link between Gaussian TDVP and SCHA.
We provided solid theoretical groundwork for the use of the SCHA dynamical ansatz in studying spectral properties.
The presence of dynamical 2-phonon excitations in Gaussian TDVP was essential to obtain correct dynamics of the 1-phonon excitations.
We compared the linearized time evolution and the projected Hamiltonian methods to find that only the former is correct in the perturbative limit.
Our work establishes a useful connection between TDVP and SCHA, allowing further developments in both fields.

\begin{acknowledgments}
This work was supported by the Creative-Pioneering Research Program through Seoul National University, Korean NRF No-2020R1A2C1014760, and the Institute for Basic Science (No. IBSR009-D1).
\end{acknowledgments}

\makeatletter\@input{xy.tex}\makeatother
\bibliography{main}

\end{document}


\title{Supplemental Material: Gaussian time-dependent variational principle for the finite-temperature anharmonic lattice dynamics}

\author{Jae-Mo Lihm}
\email{jaemo.lihm@gmail.com}
\author{Cheol-Hwan Park}
\email{cheolhwan@snu.ac.kr}
\affiliation{Center for Correlated Electron Systems, Institute for Basic Science, Seoul 08826, Korea}
\affiliation{Department of Physics and Astronomy, Seoul National University, Seoul 08826, Korea}
\affiliation{Center for Theoretical Physics, Seoul National University, Seoul 08826, Korea}

\date{\today}

\maketitle

\tableofcontents

\section{Physical meaning of the variational parameters} \label{sec:supp_transf}
In this section, we detail the physical meaning of the transformations and the tangent vectors by inspecting the infinitesimal transformation of the position and momentum operators.
Using the definition of the operator transformation~[\myeqref{eq:s_transf_op_x}] and $\left. \partial_\mu \opUd \right|_{\mbx = 0} = - \left. \partial_\mu \opU \right|_{\mbx = 0}$, one finds that the derivative of $\hat{A}(\mbx)$ at $\mbx = 0$ is
\begin{equation} \label{eq:s_transf_op_der}
    \left. \partial_\mu \hat{A}(\mbx) \right|_{\mbx = 0} = \comm{\hat{A}(0)}{\left. \partial_\mu \opU \right|_{\mbx = 0}}.
\end{equation}
Hereafter, all derivatives with respect to the variational parameters are evaluated at $\mbx = 0$ unless otherwise stated.

The derivatives of the Gaussian transformation operator at $\mbx = 0$ are
\begin{equation} \label{eq:s_overlap_dU_ar}
    \dfdx{\opU}{\alpr{m}}
    = \frac{1}{\sqrt{2}}(\opad_m - \opa_m)
    = -i\opp_m,
\end{equation}
\begin{equation} \label{eq:s_overlap_dU_ai}
    \dfdx{\opU}{\alpi{m}}
    = \frac{i}{\sqrt{2}}(\opad_m + \opa_m)
    = i\opr_m,
\end{equation}
\begin{align} \label{eq:s_overlap_dU_br}
    \dfdx{\opU}{\betr{mn}}
    = b_{mn} (\opad_m\opad_n - \opa_n\opa_m)
    = -ib_{mn} (\opr_m\opp_n + \opp_m \opr_n),
\end{align}
\begin{align} \label{eq:s_overlap_dU_bi}
    \dfdx{\opU}{\beti{mn}}
    = ib_{mn} (\opad_m\opad_n + \opa_n\opa_m)
    = ib_{mn} (\opr_m\opr_n - \opp_m \opp_n),
\end{align}
\begin{align} \label{eq:s_overlap_dU_cr}
    \dfdx{\opU}{\gamr{mn}}
    = c_{mn} (\opad_m\opa_n - \opad_n\opa_m)
    = ic_{mn} (\opr_m\opp_n - \opp_m \opr_n),
\end{align}
\begin{align} \label{eq:s_overlap_dU_ci}
    \dfdx{\opU}{\gami{mn}}
    = -i c_{mn} (\opad_m\opa_n + \opad_n\opa_m)
    = -i c_{mn} (\opr_m\opr_n + \opp_m \opp_n).
\end{align}

We calculate how the position and momentum operators transform for an infinitesimal change of each variational parameter.
For conciseness, we write the real and imaginary parts of the variational parameters as follows:
\begin{align}
    &\alpr{m}  \equiv \Re \alpha_{m}, \quad
     \betr{mn} \equiv \Re \beta_{mn}, \quad
     \gamr{mn} \equiv \Re \gamma_{mn}, \\
    &\alpi{m}  \equiv \Im \alpha_{m}, \quad
     \beti{mn} \equiv \Im \beta_{mn}, \quad
     \gami{mn} \equiv \Im \gamma_{mn}. \nonumber
\end{align}
First, for the displacement parameter $\mb{\alpha}$, the infinitesimal transformation of the position and momentum operators are
\begin{equation} \label{eq:s_transf_ar}
    \left. \dfdx{\opr_{p}(\mbx)}{\alpr{m}} \right|_{\mbx = 0} = \delta_{m,p}, \quad
    \left. \dfdx{\opp_{p}(\mbx)}{\alpr{m}} \right|_{\mbx = 0} = 0,
\end{equation}
\begin{equation} \label{eq:s_transf_ai}
    \left. \dfdx{\opr_{p}(\mbx)}{\alpi{m}} \right|_{\mbx = 0} = 0, \quad
    \left. \dfdx{\opp_{p}(\mbx)}{\alpi{m}} \right|_{\mbx = 0} = \delta_{m,p}.
\end{equation}

For the real part of the squeezing parameters $\mb{\beta}$ and $\mb{\gamma}$, the infinitesimal transformation of $\opr$ and $\opp$ are
\begin{equation} \label{eq:s_transf_br}
    \left. \dfdx{\opr_{p}(\mbx)}{\betr{mn}} \right|_{\mbx = 0}
    = b_{mn} (\opr_{m} \delta_{n,p} + \opr_{n} \delta_{m,p}), \quad
    \left. \dfdx{\opp_{p}(\mbx)}{\betr{mn}} \right|_{\mbx = 0}
    = b_{mn} (-\opp_{m} \delta_{n,p} - \opp_{n} \delta_{m,p}),
\end{equation}
\begin{equation} \label{eq:s_transf_bi}
    \left. \dfdx{\opr_{p}(\mbx)}{\gamr{mn}} \right|_{\mbx = 0}
    = c_{mn} (-\opr_{m} \delta_{n,p} + \opr_{n} \delta_{m,p}), \quad
    \left. \dfdx{\opp_{p}(\mbx)}{\gamr{mn}} \right|_{\mbx = 0}
    = c_{mn} (-\opp_{m} \delta_{n,p} + \opp_{n} \delta_{m,p}).
\end{equation}

Finally, for the imaginary part of the squeezing parameters $\mb{\beta}$ and $\mb{\gamma}$, the infinitesimal transformation of $\opr$ and $\opp$ are
\begin{equation}  \label{eq:s_transf_cr}
    \left. \dfdx{\opr_{p}(\mbx)}{\beti{mn}} \right|_{\mbx = 0}
    = b_{mn} (\opp_{m} \delta_{n,p} + \opp_{n} \delta_{m,p}), \quad
    \left. \dfdx{\opp_{p}(\mbx)}{\beti{mn}} \right|_{\mbx = 0}
    = b_{mn} (\opr_{m} \delta_{n,p} + \opr_{n} \delta_{m,p}),
\end{equation}
\begin{equation}  \label{eq:s_transf_ci}
    \left. \dfdx{\opr_{p}(\mbx)}{\gami{mn}} \right|_{\mbx = 0}
    = c_{mn} (\opp_{m} \delta_{n,p} + \opp_{n} \delta_{m,p}), \quad
    \left. \dfdx{\opp_{p}(\mbx)}{\gami{mn}} \right|_{\mbx = 0}
    = -c_{mn} (\opr_{m} \delta_{n,p} + \opr_{n} \delta_{m,p}).
\end{equation}

From Eqs.~(\ref{eq:s_transf_ar}-\ref{eq:s_transf_ci}), one can understand the role of each variational parameter.
The real part of the displacement parameter, $\alpr{m}$, parametrizes the displacement of the position operator for mode $m$. These $N$ degrees of freedom corresponds to the center position $\widetilde{R}$ in the SCHA harmonic Hamiltonian.
The real parts of the squeezing parameters, $\betr{mn}$ and $\gamr{mn}$, parametrize the change in the normal mode frequency and eigenvectors.
Especially, $\gamr{mn}$ parametrizes the linear combination of the two eigenmodes $m$ and $n$.

If modes $m$ and $n$ are nondegenerate, setting $\gamr{mn} \neq 0$ mixes two modes with different frequencies, inducing a nontrivial transformation of the thermal density matrix.
In contrast, if modes $m$ and $n$ are degenerate (i.e. $\omega_m = \omega_n$), the linear combination parametrized by $\gamr{mn}$ is a gauge transformation that does not change the density matrix.
Hence, it is justified to exclude $\gamr{mn}$ from the variational parameters when modes $m$ and $n$ are degenerate, as mentioned in the main text.
From a theoretical point of view, including $\gamma_{mn}$ in the set of variational parameters for degenerate modes $m$ and $n$ makes the symplectic form [\myeqref{eq:tdvp_Omega_def} of the main text] noninvertible and thus should be avoided~\cite{2020HacklReview}.

The imaginary parts of the Gaussian parameters generate dynamics of the variational states.
The displacement parameter $\alpi{m}$ parametrizes the generation of finite atomic momentum.
The squeezing parameters $\beti{mn}$ and $\gami{mn}$ parametrize the linear combination of the position coordinates with the momentum coordinates and vice versa.

\section{Linear response formulation of TDVP at finite temperatures} \label{sec:supp_tdvp}
In this section, we derive and summarize the key results of the linear response formulation of TDVP at finite temperatures, following Ref.~\cite{2020HacklReview}.

In \myeqref{eq:pur_psi_var} of the main text, we mapped the variational density matrices to pure state wavefunctions by purification.
The maximally entangled state $\ket{\Phi^+}$ is defined as 
\begin{equation} \label{eq:s_tdvp_Phi_def}
    \ket{\Phi^+} \sim \sum_{n_1,\cdots,n_{N}} \ket{n_1, \cdots, n_{N}} \otimes \ket{n_1, \cdots, n_{N}}.
\end{equation}
Thanks to the unitarity of $\opU$, the variational wavefunction $\ket{\Psi(\mbx)}$ is always normalized to unity.
The original density matrix is recovered by taking a partial trace of the auxiliary system:
\begin{equation} \label{eq:s_tdvp_psi_tr}
    \oprho(\mbx) = \Tr_{\rm aux} \ket{\Psi(\mbx)} \bra{\Psi(\mbx)}.
\end{equation}

Note that although $\ket{\Psi(0)}$ is a purification of the thermal state $\oprho_0$ of the harmonic Hamiltonian $\opH^{\rm (H)}$, it is not a stationary state of the time evolution with $\opH^{\rm (H)}$:
\begin{align} \label{eq:s_tdvp_psi_time}
    (e^{-i\opH^{\rm (H)} t} \otimes \one) \ket{\Psi(0)}
    =& (\sqrt{\oprho_0} e^{-i\opH^{\rm (H)} t} \otimes \one) \ket{\Phi^+} \nnnl
    =& (\sqrt{\oprho_0} \otimes e^{-i\opH^{\rm (H)} t}) \ket{\Phi^+} \nnnl
    \neq& \ket{\Psi(0)} e^{i\phi(t)}
\end{align}
for any choice of the phase $\phi(t)$.
In the second equality of \myeqref{eq:s_tdvp_psi_time}, we used the fact that $\opH^{\rm (H)}$ is diagonal in the eigenmode basis.
Still, the corresponding density matrix that is obtained by taking the partial trace of the auxiliary system is time-independent.
Hence, the time evolution of the purified wavefunction is not a true dynamics in the physical system.
It is an auxiliary dynamics that occurs due to the non-uniqueness of the purification up to a unitary transformation at the auxiliary system.
This artificial dynamics does not occur in our variational approach because we do not allow any variational degree of freedom to the auxiliary system.

The time evolution of the variational wavefunction is obtained by projecting the change in the wavefunction to the tangent space of the variational manifold.
The tangent space is spanned by the tangent vectors, which are the derivatives of the variational wavefunction orthogonalized to the original wavefunction.
Formally, the tangent vectors are defined as
\begin{align} \label{eq:s_tdvp_tan_def_1}
    \ket{V_\mu(\mbx)}
    =& \hat{Q}(\mbx) \left. \dfdx{\ket{\Psi(\mbx)}}{x^\mu} \right|_{\mbx}.
\end{align}
where $\hat{Q}(\mbx)$ a projection operator:
\begin{equation}
    \hat{Q}(\mbx)
    = \one - \ket{\Psi(\mbx)} \bra{\Psi(\mbx)}.
\end{equation}

According to TDVP, the dynamics of the variational parameters can be described by a classical Hamilton equation of motion.
To determine the equation of motion, we need the symplectic form and the derivatives of the energy expectation value $E(\mbx) = \mel{\Psi(\mbx)}{\opH}{\Psi(\mbx)}$~\cite{2020HacklReview}.

The symplectic form $\Omega^{\mu\nu}(\mb{x})$ is the inverse of $\omega_{\mu\nu}(\mb{x})$, which is twice the imaginary part of the inner product of the tangent vectors:
\begin{equation} \label{eq:s_tdvp_Omega_def}
    \Omega^{\mu \rho}(\mb{x}) \omega_{\rho \nu}(\mb{x}) = \delta^{\mu}_{\hpmu\nu},
\end{equation}
\begin{equation} \label{eq:s_tdvp_omega_def}
\omega_{\mu\nu}(\mb{x}) = 2 \Im \braket{V_\mu(\mb{x})}{V_\nu(\mb{x})}.
\end{equation}
We use Greek indices to denote the components of the real-valued vector $\mbx$ defined in \myeqref{eq:tdvp_x_def} of the main text.
We use Einstein's summation convention for repeated indices.

According to the Lagrangian action principle, the equation of motion of the variational parameters is~\cite{2018ShiNonGaussian, 2020HacklReview}
\begin{equation} \label{eq:s_tdvp_eom}
    \frac{dx^\mu}{dt}
    = -\Omega^{\mu \nu}(\mb{x}) \left. \dfdx{E(\mbx)}{x^\nu} \right|_{\mbx}.
\end{equation}
We note that since the Gaussian variational manifold is a K\"ahler mainfold, the Lagrangian, McLachlan, and Dirac-Frenkel TDVP equations are all equivalent~\cite{2020HacklReview}.

Now, we illustrate how to compute dynamical and spectral properties using the linear response formulation of TDVP.
As we are interested only in small changes of the wavefunction around the stationary state, we linearize the equation of motion \myeqref{eq:s_tdvp_eom} around $\mbx = 0$ to find~\cite{2018ShiNonGaussian,2019GuitaBosonic,2020HacklReview}
\begin{equation} \label{eq:s_tdvp_eom_lin}
    \frac{dx^\mu}{dt} = K^{\mu}_{\hpmu \nu} x^\nu,
\end{equation}
where the linearized time-evolution generator $\mb{K}$ is
\begin{align} \label{eq:s_tdvp_K_def}
K^\mu_{\hpmu\nu}
= \left. \frac{\partial}{\partial x^\nu} \left( -\Omega^{\mu\rho}(\mbx) \dfdx{E}{x^\rho} \right) \right|_{\mbx = 0}
= -\Omega^{\mu\rho}(\mbx=0) \left. \dfdxy{E}{x^\rho}{x^\nu} \right|_{\mbx = 0},
\end{align}
as shown in \myeqref{eq:tdvp_K_def} of the main text.
Here, we used $\left. (\partial E / \partial x^\rho) \right|_{\mbx = 0}=0$ which is true because $\mbx = 0$ is a stationary point.
From now on, we denote $\partial / \partial x^\mu$ by $\partial_\mu$.
Also, we use $\Omega^{\mu\rho}$ to refer to $\Omega^{\mu\rho}(\mbx = 0)$ unless otherwise noted.
The solution of the linearized equation of motion is
\begin{equation} \label{eq:s_tdvp_lin_sol}
    x^\mu(t) = [\mb{d\Phi}(t)]^{\mu}_{\hpmu \nu} x^\nu(0),
\end{equation}
where $\mb{d\Phi}(t)$ is the linearized free evolution flow defined as
\begin{equation} \label{eq:s_lr_Phi_def}
    \mb{d\Phi}(t) = e^{\mb{K}t}.
\end{equation}

Let us consider a standard linear response setting, where an infinitesimal time-dependent perturbation is added to the Hamiltonian:
\begin{equation}
    \hat{H}_{\epsilon}(t) = \hat{H} + \epsilon \varphi(t)\hat{A}.
\end{equation}
Here, $\hat{A}$ is an arbitrary Hermitian operator in the Hilbert space of purified wavefunctions, $\varphi(t)$ is a real-valued function, and $\epsilon$ is a real variable parametrizing the strength of the perturbation.
We write the solution of the corresponding variational time evolution as $\ket{\Psi_\epsilon (t)} \equiv \ket{\Psi(\mbx_\epsilon (t))}$.

The linear response of the variational parameter is defined as
\begin{equation}
    \delta_A x^\mu (t) = \left. \frac{d}{d \epsilon} x^\mu_\epsilon (t) \right|_{\epsilon = 0}.
\end{equation}
According to Proposition 8 of Ref.~\cite{2020HacklReview}, $\delta_A x^\mu (t)$ is given as
\begin{equation} \label{eq:s_lr_dxt}
    \delta_A x^\mu (t) = - \Omega^{\nu\rho} \partial_\rho A \int_{-\infty}^{t} dt' [\mb{d\Phi}(t-t')]^\mu_{\hpmu\nu} \varphi(t'),
\end{equation}
where
\begin{equation} \label{eq:s_lr_dA_def}
    \partial_\rho A
    \equiv \left. \frac{\partial}{\partial x^\rho} \mel{\Psi(\mbx)}{\hat{A}}{\Psi(\mbx)} \right|_{\mbx = 0}.
\end{equation}
The linear response of the expectation value of an operator $\hat{B}$ at time $t$ is~\cite{2020HacklReview}
\begin{align} \label{eq:s_lr_dB}
    \delta_A B(t)
    =& \left. \frac{d}{d\epsilon} \mel{\Psi_\epsilon (t)}{\hat{B}}{\Psi_\epsilon (t)} \right|_{\epsilon=0} \nnnl
    =& \delta_A x^\mu (t) \partial_\mu B \\
    =& - (\partial_\mu B) (\Omega^{\nu\rho} \partial_\rho A) \int_{-\infty}^{t} dt' [\mb{d\Phi}(t-t')]^\mu_{\hpmu\nu} \varphi(t'). \nonumber
\end{align}

Now, we use the spectral decomposition of $\mb{K}$ to compute $\mb{d\Phi}(t)$.
One can decompose $i\mb{K}$ with eigenvalues $\lambda_l$, eigenvectors $\mcE^\mu(\lambda_l)$ and dual eigenvectors $\overline{\mcE_\nu}(\lambda_l)$~\cite{2020HacklReview}:
\begin{equation} \label{eq:s_lr_iK_eigen}
    iK^\mu_{\hpmu\nu} = \sum_{l} \lambda_l \mcE^\mu(\lambda_l) \overline{\mcE_\nu}(\lambda_l).
\end{equation}
The dual eigenvectors satisfy
\begin{equation} \label{eq:s_lr_duale_def}
    \overline{\mcE_\mu}(\lambda_l) \mcE^\mu(\lambda_{l'}) = \delta_{l,l'}.
\end{equation}
Then, the linearized free evolution flow becomes
\begin{equation} \label{eq:s_lr_dphi}
    [\mb{d\Phi}(t)]^\mu_{\hpmu\nu} = \sum_{l} e^{-i\lambda_l t} \mcE^\mu(\lambda_l) \overline{\mcE_\nu}(\lambda_l).
\end{equation}

Using Eq.~\eqref{eq:s_lr_dB} and Eq.~\eqref{eq:s_lr_dphi}, we find
\begin{align} \label{eq:s_lr_dB2}
    \delta_{A} B(t)
    =& - \sum_{l} [\mcE^\mu(\lambda_l) \partial_\mu B] [\overline{\mcE_\nu}(\lambda_l) \Omega^{\nu\rho} \partial_\rho A]
    \int_{-\infty}^{t} dt' e^{-i\lambda_l (t-t')} \varphi(t').
\end{align}
By taking the Fourier transform of Eq.~\eqref{eq:s_lr_dB2}, we find
\begin{align} \label{eq:s_lr_dBw}
    \delta_A B(\omega)
    =& -i \varphi(\omega) \sum_{l} [\mcE^\mu(\lambda_l) \partial_\mu B] [\overline{\mcE_\nu}(\lambda_l) \Omega^{\nu\rho} \partial_\rho A] 
    \lim_{\eta \rightarrow 0^+} \frac{1}{\omega - \lambda_l + i\eta}.
\end{align}
The retarded correlation function $G_{AB}^{\rm (R)}(\omega)$ is defined as
\begin{equation} \label{eq:s_lr_GAB_def}
    \delta_A B(\omega) = G_{AB}^{\rm (R)}(\omega) \varphi(\omega).
\end{equation}
From Eqs.~\eqref{eq:s_lr_GAB_def} and \eqref{eq:s_lr_dBw}, we find
\begin{equation} \label{eq:s_lr_GAB_result}
    G_{AB}^{\rm (R)}(\omega)
    = \lim_{\eta \rightarrow 0^+} -i \sum_{l} \frac{[\mcE^\mu(\lambda_l) \partial_\mu B] [\overline{\mcE_\nu}(\lambda_l) \Omega^{\nu\rho} \partial_\rho A]}{\omega + i\eta - \lambda_l}.
\end{equation}
Then, using the definition of the matrix Green function [\myeqref{eq:tdvp_G_def}], we find \myeqref{eq:tdvp_lr_eqn} of the main text.

\section{Derivation of the symplectic form} \label{sec:supp_overlap}
In this section, we calculate the overlap of the tangent vectors to calculate the metric and the symplectic form.
From the definition of the tangent vectors [\myeqref{eq:tdvp_tan_def_2}], one finds
\begin{align} \label{eq:s_overlap_vv}
    \braket{V_\mu}{V_\nu}
    = \mel{\Phi^+}{ \left( \sqrt{\oprho_0} \dfdx{\hat{U}^\dagger}{x^\mu} \dfdx{\hat{U}}{x^\nu} \sqrt{\oprho_0} \otimes \one \right) }{\Phi^+}
    = \expval{ \dfdx{\hat{U}^\dagger}{x^\mu} \dfdx{\hat{U}}{x^\nu} }_0.
\end{align}
To evaluate \myeqref{eq:s_overlap_vv}, we use the derivatives of the variational transformation, Eqs.~(\ref{eq:s_overlap_dU_ar}-\ref{eq:s_overlap_dU_ci}).

Now, we compute the overlap.
First, since the thermal expectation value of an operator containing uneven numbers of creation and annihilation operators is zero, one finds
\begin{equation} \label{eq:s_overlap_a_bc}
    \expval{\dfdx{\hat{U}^\dagger}{\alpha^{\rm r/i}_{m}} \dfdx{\hat{U}}{\beta^{\rm r/i}_{pq}}}_0
    = \expval{\dfdx{\hat{U}^\dagger}{\alpha^{\rm r/i}_{m}} \dfdx{\hat{U}}{\gamma^{\rm r/i}_{pq}}}_0
    = 0.
\end{equation}
and
\begin{equation} \label{eq:s_overlap_b_c}
    \expval{\dfdx{\hat{U}^\dagger}{\beta^{\rm r/i}_{mn}} \dfdx{\hat{U}}{\gamma^{\rm r/i}_{pq}}}_0
    = 0.
\end{equation}

Next, we calculate the nonzero inner products.
First, for two displacement parameters $\alpha_m$ and $\alpha_n$, we find
\begin{equation} \label{eq:s_overlap_ar_ar}
    \expval{\dfdx{\hat{U}^\dagger}{\alpr{m}} \dfdx{\hat{U}}{\alpr{n}}}_0
    = \expval{\opp_m \opp_n }_0
    = \left( n_m + \frac{1}{2} \right) \delta_{m,n},
\end{equation}
\begin{equation} \label{eq:s_overlap_ar_ai}
    \expval{\dfdx{\hat{U}^\dagger}{\alpr{m}} \dfdx{\hat{U}}{\alpi{n}}}_0
    = -\expval{\opp_m \opr_n}_0
    = \frac{i}{2} \delta_{m,n},
\end{equation}
and
\begin{equation} \label{eq:s_overlap_ai_ai}
    \expval{\dfdx{\hat{U}^\dagger}{\alpi{m}} \dfdx{\hat{U}}{\alpi{n}}}_0
    = \expval{\opr_m \opr_n}_0
    = \left( n_m + \frac{1}{2} \right) \delta_{m,n}.
\end{equation}

Next, we consider the tangent vectors of the squeezing parameters $\betr{mn}$ and $\betr{pq}$.
Note that
\begin{equation}
    \delta_{m,p}\delta_{n,q} + \delta_{m,q}\delta_{n,p}
    = \begin{cases}
    \delta_{mn, pq} & \text{if $m \neq n$} \\
    2 \delta_{mn, pq} & \text{if $m = n$} \\
    \end{cases}
\end{equation}
holds since $m \leq n$ and $p \leq q$.
Then, using Eqs.~(\ref{eq:s_overlap_dU_br}, \ref{eq:s_overlap_dU_bi}), we find
\begin{align} \label{eq:s_overlap_br_br}
    \expval{\dfdx{\hat{U}^\dagger}{\betr{mn}} \dfdx{\hat{U}}{\betr{pq}}}_0
    = \expval{\dfdx{\hat{U}^\dagger}{\beti{mn}} \dfdx{\hat{U}}{\beti{pq}}}_0
    =& b_{mn}b_{pq} \expval{\opa_n \opa_m \opad_p \opad_q + \opad_m \opad_n \opa_q \opa_p}_0 \nnnl
    =& b_{mn}b_{pq} (\delta_{m,p}\delta_{n,q} + \delta_{m,q}\delta_{n,p})
    (2 n_m n_n + n_m + n_n + 1) \nnnl
    =& \frac{2 n_m n_n + n_m + n_n + 1}{2(n_m + n_n + 1)} \delta_{mn, pq},
\end{align}
and
\begin{align}\label{eq:s_overlap_br_bi}
    \expval{\dfdx{\hat{U}^\dagger}{\betr{mn}} \dfdx{\hat{U}}{\beti{pq}}}_0
    =& i b_{mn}b_{pq} \expval{\opa_n \opa_m \opad_p \opad_q - \opad_m \opad_n \opa_q \opa_p}_0 \nnnl
    =& i b_{mn}b_{pq} (\delta_{m,p}\delta_{n,q} + \delta_{m,q}\delta_{n,p})
    (n_m + n_n + 1) \nnnl
    =& \frac{i}{2} \delta_{mn, pq}.
\end{align}

Finally, for the tangent vectors of the squeezing parameters $\gamr{mn}$ and $\gamr{pq}$, $m < n$ and $p < q$ holds by definition. Thus, we find
\begin{align} \label{eq:s_overlap_cr_cr}
    \expval{\dfdx{\hat{U}^\dagger}{\gamr{mn}} \dfdx{\hat{U}}{\gamr{pq}}}_0
    = \expval{\dfdx{\hat{U}^\dagger}{\gami{mn}} \dfdx{\hat{U}}{\gami{pq}}}_0
    =& c_{mn}c_{pq} \expval{(\opad_n \opa_m - \opad_m \opa_n)(\opad_p \opa_q - \opad_q \opa_p)}_0 \nnnl
    =& c_{mn}c_{pq} \delta_{m,p}\delta_{n,q}
    (2 n_m n_n + n_m + n_n) \nnnl
    =& \frac{2 n_m n_n + n_m + n_n}{2(n_m - n_n)} \delta_{mn, pq},
\end{align}
and
\begin{align} \label{eq:s_overlap_cr_ci}
    \expval{\dfdx{\hat{U}^\dagger}{\gamr{mn}} \dfdx{\hat{U}}{\gami{pq}}}_0
    =& -i c_{mn}c_{pq} \expval{(\opad_n \opa_m - \opad_m \opa_n)(\opad_p \opa_q + \opad_q \opa_p)}_0 \nnnl
    =& -i c_{mn}c_{pq} \delta_{m,p}\delta_{n,q}
    (n_n - n_m) \nnnl
    =& \frac{i}{2} \delta_{mn,pq}.
\end{align}

The only inner products with nonzero imaginary parts are those in Eqs.~(\ref{eq:s_overlap_ar_ai}, \ref{eq:s_overlap_br_bi}, \ref{eq:s_overlap_cr_ci}) and their complex conjugates.
Using this result and the definition of the symplectic form [\myeqref{eq:tdvp_Omega_def}], one obtains \myeqref{eq:anh_Omega} of the main text.

\section{Derivatives of the energy} \label{sec:supp_deriv}
In this section, we calculate the first and second derivatives of the energy expectation value with respect to the variational parameters.
In this section, all derivatives are evaluated at $\mb{x}=\mb{0}$ unless otherwise noted.

\subsection{Useful identities}
Before actually calculating the derivatives, we derive useful identities.
Using the normal mode representation of the anharmonic Hamiltonian [\myeqref{eq:scha_ham_mode_def}], we find
\begin{equation} \label{eq:s_der_iden_Hr_comm}
    \comm{\opH}{\opr_m} = -i \omega_m \opp_m,
\end{equation}
and
\begin{equation} \label{eq:s_der_iden_Hp_comm}
    \comm{\opH}{\opp_m} = i \dfdx{\opV}{r_m}.
\end{equation}
Also, given an observable $\hat{O} = O(\hat{\mb{r}})$ which is a function of the position operators, one finds
\begin{align} \label{eq:s_der_iden_rO}
    \expval{\opr_m \hat{O}}_0
    = \int d\mb{r} \rho_0(\mb{r}) \opr_m O(\mb{r})
    = - \left( n_m + \frac{1}{2} \right) \int d\mb{r} \dfdx{\rho_0(\mb{r})}{r_m} O(\mb{r})
    = \left( n_m + \frac{1}{2} \right) \int d\mb{r} \rho_0(\mb{r}) \dfdx{O(\mb{r})}{r_m}
    = \left( n_m + \frac{1}{2} \right) \expval{\dfdx{\hat{O}}{r_m}}_0.
\end{align}
Here, $\rho_0(\mb{r})$ is the diagonal part of $\oprho_0$ in the normal mode position basis~\cite{PathriaBeale_StatMech}:
\begin{equation} \label{eq:s_der_rho_0_r}
    \rho_0(\mb{r}) = \mel{\mb{r}}{\oprho_0}{\mb{r}}
    = \prod_{m=1}^{N} \sqrt{\frac{1}{\pi(2n_m + 1)}} \exp(- \frac{r_m^2}{2n_m + 1}).
\end{equation}
In the third equality of \myeqref{eq:s_der_iden_rO}, we used a partial integration with respect to $r_m$ [see also Eqs.~(C1-C3) of Ref.~\cite{2017Bianco}].

In addition, using
\begin{equation}
    e^{-\beta \hat{H}^\mathrm{(H)}} \opa_m
    = e^{\beta \omega_m} \opa_m e^{-\beta \hat{H}^\mathrm{(H)}},
\end{equation}
one can show
\begin{equation} \label{eq:s_der_iden_aO}
    \expval{\opa_m \hat{O}}_0 = e^{\beta \omega_m} \expval{\hat{O} \opa_m}_0
\end{equation}
and
\begin{equation} \label{eq:s_der_iden_adO}
    \expval{\opad_m \hat{O}}_0 = e^{-\beta \omega_m} \expval{\hat{O} \opad_m}_0.
\end{equation}
From Eqs.~\eqref{eq:s_der_iden_aO} and \eqref{eq:s_der_iden_adO}, one can show
\begin{equation} \label{eq:s_der_iden_pO_derivation}
    \expval{\hat{O} \opr_m}_0
    = \frac{e^{\beta \omega_m}}{\sqrt{2}} \expval{\opad_m \hat{O}}_0
    + \frac{e^{-\beta \omega_m}}{\sqrt{2}} \expval{\opa_m \hat{O}}_0
    = \frac{e^{\beta \omega_m} + e^{-\beta \omega_m}}{2} \expval{\opr_m \hat{O}}_0
    + \frac{e^{\beta \omega_m} - e^{-\beta \omega_m}}{2i} \expval{\opp_m \hat{O}}_0.
\end{equation}
Using \myeqref{eq:s_der_iden_pO_derivation} and $\comm{\opr_m}{O(\hat{\mb{r}})} = 0$, one finds
\begin{equation} \label{eq:s_der_iden_pO}
    \expval{\opp_m \hat{O}}_0
    = -\frac{i}{2 n_m + 1} \expval{\opr_m \hat{O}}_0
    = -\frac{i}{2} \expval{\dfdx{\hat{O}}{r_m}}_0.
\end{equation}
Taking complex conjugate of \myeqref{eq:s_der_iden_pO}, one also finds
\begin{equation} \label{eq:s_der_iden_Op}
    \expval{\hat{O} \opp_m}_0
    = \frac{i}{2} \expval{\dfdx{\hat{O}}{r_m}}_0.
\end{equation}
Using a logic similar to \myeqref{eq:s_der_iden_pO_derivation}, one can also show
\begin{align} \label{eq:s_der_iden_ppO}
    \expval{\opp_m \hat{O} \opp_n}_0
    =& -\frac{i}{2 n_m + 1} \expval{\opr_m \hat{O} \opp_n}_0
    + \frac{2n_m (n_m+1)}{2n_m + 1} \delta_{m,n} \expval{\hat{O}}_0 \nnnl
    =& \frac{1}{2(2 n_m + 1)} \left( \delta_{m,n} \expval{\hat{O}}_0
    + \expval{\opr_m \dfdx{\hat{O}}{r_n}}_0 \right)
    + \frac{2n_m (n_m+1)}{2n_m + 1} \delta_{m,n} \expval{\hat{O}}_0 \nnnl
    =& \frac{1}{4} \expval{\dfdxy{\hat{O}}{r_m}{r_n}}_0
    + \left( n_m + \frac{1}{2} \right) \delta_{m,n} \expval{\hat{O}}_0.
\end{align}
We use \myeqref{eq:s_der_iden_ppO} only in \myeqref{eq:s_der_2nd_comm_pp_pp}.

\subsection{First derivatives}
Now, we compute the first derivatives of the energy expectation value and show that the SCHA solution is also the stationary state of the Gaussian TDVP.
By setting $\hat{O} = \opH$ in \myeqref{eq:s_transf_op_der} and taking the equilibrium expectation value, the first-order derivative of the energy expectation value becomes
\begin{equation} \label{eq:s_der_dE}
    \dfdx{E}{x^\mu}
    = \expval{ \comm{\opH}{\dfdx{\opU}{x^\mu}} }_0.
\end{equation}
So, the first-order derivatives can be computed using the derivatives of the Gaussian transformation operator, Eqs.~(\ref{eq:s_overlap_dU_ar}-\ref{eq:s_overlap_dU_ci}).

Using the identities [Eqs.~(\ref{eq:s_der_iden_Hr_comm}-\ref{eq:s_der_iden_Op})] as well as the properties of the SCHA density matrix [Eqs.~(\ref{eq:scha_dv}, \ref{eq:scha_p})], the first-order derivatives of energy at $\mb{x} = \mb{0}$ can be computed as follows.
We find that all first-order derivatives of the energy are zero.
For the variational parameters included in the SCHA theory, the centroid position and the force constants, the stationarity of $\oprho_0$ is expected since $\oprho_0$ is the variational solution that minimizes the SCHA free energy.
$\oprho_0$ is also stationary with respect to the variation of other parameters such as the atomic momentum parameter $\alpi{m}$ because it is a thermal density matrix whose momentum expectation value is zero.

\begin{equation}
    \dfdx{E}{\alpr{m}}
    = \expval{\left[ \opH, -i \opp_m \right]}_0
    = \expval{ \dfdx{\hat{V}}{r_m}}_0
    = 0
\end{equation}
\begin{equation}
    \dfdx{E}{\alpi{m}}
    = \expval{\left[ \opH, i \opr_m \right]}_0
    = \omega_m \expval{ \opp_m }_0
    = 0
\end{equation}

\begin{align} \label{eq:s_der_dE_betr}
    \dfdx{E}{\betr{mn}}
    =& -ib_{mn} \expval{ \comm{\opH}{\opr_m\opp_n + \opp_m\opr_n} }_0 \nnnl
    =& -ib_{mn} \Big[ -i\omega_m \expval{\opp_m\opp_n}_0 + i \expval{\opr_m \dfdx{\opV}{r_n}}_0 + (n\leftrightarrow m) \Big] \nnnl
    =& b_{mn} \Big[ -\omega_m \left( n_m + \frac{1}{2} \right) \delta_{m,n} + \left( n_m + \frac{1}{2} \right) \expval{\dfdxy{\opV}{r_m}{r_n}}_0
    + (n\leftrightarrow m) \Big] \nnnl
    =& 0
\end{align}
\begin{align} \label{eq:s_der_dE_beti}
    \dfdx{E}{\beti{mn}}
    =& ib_{mn} \expval{\left[ \opH, \opr_m \opr_n - \opp_m \opp_n \right]}_0 \nnnl
    =& i b_{mn} \Big[ -i\omega_m \expval{\opp_m \opr_n} -i\omega_n \expval{\opr_m \opp_n}
    -i\expval{\opp_m \dfdx{\opV}{r_n}}_0 - i \expval{\dfdx{\opV}{r_m} \opp_n}_0 \Big] \nnnl
    =& i b_{mn} \left[ -\frac{1}{2} \expval{\dfdxy{\opV}{r_m}{r_n}}_0 + \frac{1}{2} \expval{\dfdxy{\opV}{r_m}{r_n} }_0 \right] \nnnl
    =& 0
\end{align}

\begin{align} \label{eq:s_der_dE_gamr}
    \dfdx{E}{\gamr{mn}}
    = ic_{mn} \expval{ \comm{\opH}{\opr_m\opp_n - \opp_m\opr_n} }_0
    = 0
\end{align}
\begin{align} \label{eq:s_der_dE_gami}
    \dfdx{E}{\gami{mn}}
    = -ic_{mn} \expval{\left[ \opH, \opr_m\opr_n + \opp_m\opp_n \right]}_0
    = 0
\end{align}
Equations~\eqref{eq:s_der_dE_gamr} and \eqref{eq:s_der_dE_gami} can be derived in the same way as Eqs.~\eqref{eq:s_der_dE_betr} and \eqref{eq:s_der_dE_beti}, respectively.

\subsection{Second derivatives} \label{sec:supp_der_2}
Next, we calculate the second derivatives of energy. The result of this subsection can be summarized in a matrix form:
\begin{equation} \label{eq:s_der_ddE_mat}
\dfdxy{E}{x^\mu}{x^\nu}
= \begin{pmatrix}
\mb{\omega} & 0 & \mbPht B & 0 & \mbPht C & 0\\
0 & \mb{\omega} & 0 & 0 & 0 & 0\\
B \mbPht & 0 & \mbom_{+} + B \mbPhf  B & 0 & B\mbPhf C & 0\\
0 & 0 & 0 & \mbom_{+} & 0 & 0 \\
C \mbPht & 0 & C \mbPhf  B & 0 & \mbom_{-} + C \mbPhf  C & 0\\
0 & 0 & 0 & 0 & 0 & \mbom_{-}
\end{pmatrix}.
\end{equation}

The remaining part of this subsection is the derivation of \myeqref{eq:s_der_ddE_mat}.
By taking derivative of \myeqref{eq:s_transf_op_x} with $\hat{O} = \opH$,
the second derivative of energy at $\mbx = 0$ is given by
\begin{align} \label{eq:s_der_ddU}
    \dfdxy{E}{x^\mu}{x^\nu}
    = \Bigg< \opH \dfdxy{\opU}{x^\mu}{x^\nu}
    + \dfdxy{\opUd}{x^\mu}{x^\nu} \opH
    + \dfdx{\opUd}{x^\mu} \opH \dfdx{\opU}{x^\nu}
    + \dfdx{\opUd}{x^\nu} \opH \dfdx{\opU}{x^\mu} \Bigg>_0.
\end{align}

When the two derivatives are for the same parameter type (displacement or squeezing), the second derivative of the transformation matrix becomes
\begin{equation} \label{eq:s_der_ddU_1}
    \dfdxy{\opU}{x^\mu}{x^\nu}
    = \frac{1}{2} \acomm{\dfdx{\opU}{x^\mu}}{\dfdx{\opU}{x^\nu}}.
\end{equation}
In this case, using $\partial_\mu \opUd = -\partial_\mu \opU$, the second derivative of the energy can be written as
\begin{align} \label{eq:s_der_ddE_1}
    \dfdxy{E}{x^\mu}{x^\nu}
    =& \frac{1}{2} \Bigg< \opH \acomm{\dfdx{\opU}{x^\mu}}{\dfdx{\opU}{x^\nu}} 
    + \acomm{\dfdx{\opU}{x^\mu}}{\dfdx{\opU}{x^\nu}} \opH
    - 2\dfdx{\opU}{x^\mu} \opH \dfdx{\opU}{x^\nu}
    - 2\dfdx{\opU}{x^\nu} \opH \dfdx{\opU}{x^\mu} \Bigg>_0 \nnnl
    =& \frac{1}{2} \expval{ \comm{ \comm{\opH}{\dfdx{\opU}{x^\mu}} }{\dfdx{\opU}{x^\nu}} }_0
    + (\mu \leftrightarrow \nu).
\end{align}

For mixed second derivatives in which the derivatives are with respect to one displacement and one squeezing parameter, one finds
\begin{equation} \label{eq:s_der_ddU_2}
    \dfdxy{\opU}{\alpha^{\rm r/i}_p}{\beta^{\rm r/i}_{mn}}
    = \dfdx{\opU}{\alpha^{\rm r/i}_p}  \dfdx{\opU}{\beta^{\rm r/i}_{mn}},
\end{equation}
and the same for $\gamma$ instead of $\beta$.
In this case, the second derivative of energy becomes
\begin{align} \label{eq:s_der_ddE_2}
    \dfdxy{E}{\alpha^{\rm r/i}_p}{\beta^{\rm r/i}_{mn}}
    =& \Bigg< \opH \dfdx{\opU}{\alpha^{\rm r/i}_p} \dfdx{\opU}{\beta^{\rm r/i}_{mn}}
    + \dfdx{\opU}{\beta^{\rm r/i}_{mn}} \dfdx{\opU}{\alpha^{\rm r/i}_p} \opH
    - \dfdx{\opU}{\alpha^{\rm r/i}_p} \opH \dfdx{\opU}{\beta^{\rm r/i}_{mn}}
    - \dfdx{\opU}{\beta^{\rm r/i}_{mn}} \opH \dfdx{\opU}{\alpha^{\rm r/i}_p} \Bigg>_0 \nnnl
    =& \expval{ \comm{ \comm{\opH}{\dfdx{\opU}{\alpha^{\rm r/i}_p}} }{\dfdx{\opU}{\beta^{\rm r/i}_{mn}}} }_0,
\end{align}
and the same for $\gamma$ instead of $\beta$.

For the second derivatives with respect to two displacement parameters $\alpha_m$ and $\alpha_n$, we use \myeqref{eq:s_der_ddE_1} to find
\begin{align}
    \dfdxy{E}{\alpr{m}}{\alpr{n}}
    = -\frac{1}{2} \expval{ \comm{ \comm{\opH}{\opp_m} }{\opp_n} }_0 + (m \leftrightarrow n)
    = -\frac{i}{2} \expval{ \comm{ \dfdx{\opV}{r_m} }{\opp_n} }_0  + (m \leftrightarrow n)
    = \expval{ \dfdxy{\opV}{r_m}{r_n} }_0
    =& \omega_m \delta_{m,n},
\end{align}
\begin{align}
    \dfdxy{E}{\alpr{m}}{\alpi{n}}
    = \frac{1}{2} \expval{ \comm{ \comm{\opH}{\opp_m} }{\opr_n} }_0
    + \frac{1}{2} \expval{ \comm{ \comm{\opH}{\opr_n} }{\opp_m} }_0
    = \frac{i}{2} \expval{ \comm{ \dfdx{\opV}{r_m} }{\opr_n} }_0
    - \frac{i}{2} \omega_n \expval{ \comm{ \opp_n }{\opp_m} }_0
    =& 0,
\end{align}
and
\begin{align}
    \dfdxy{E}{\alpi{m}}{\alpi{n}}
    = -\frac{1}{2} \expval{ \comm{ \comm{\opH}{\opr_m} }{\opr_n} }_0 + (m \leftrightarrow n)
    = \frac{i}{2} \omega_m \expval{ \comm{ \opp_m }{\opr_n} }_0 + (m \leftrightarrow n)
    =& \omega_m \delta_{m,n}.
\end{align}

Similarly, one can also calculate the second derivatives with respect to two squeezing parameters.
Before going on, we first list some useful identities related to nested commutators.

\begin{align}
    \comm{\opH}{\opr_m\opr_n}
    = \opr_m \comm{\opH}{\opr_n} + \comm{\opH}{\opr_m} \opr_n
    = -i (\omega_n \opr_m \opp_n + \omega_m \opp_m \opr_n)
\end{align}

\begin{align}
    \comm{\opH}{\opr_m\opp_n}
    = \opr_m \comm{\opH}{\opp_n} + \comm{\opH}{\opr_m} \opp_n
    = i \opr_m \dfdx{\opV}{r_n} - i\omega_m \opp_m \opp_n
\end{align}

\begin{align}
    \comm{\opH}{\opp_m\opp_n}
    = \opp_m \comm{\opH}{\opp_n} + \comm{\opH}{\opp_m} \opp_n
    = i \opp_m \dfdx{\opV}{r_n} + i \dfdx{\opV}{r_m} \opp_n
\end{align}

\begin{align} \label{eq:s_der_2nd_comm_xx_xx}
    \expval{ \comm{ \comm{\opH}{\opr_m\opr_n} }{ \opr_p \opr_q } }_0
    =& \expval{ \comm{ -i (\omega_n \opr_m \opp_n + \omega_m \opp_m \opr_n) }{ \opr_p \opr_q } }_0 \nnnl
    =& -i \omega_n \expval{ \opr_m \comm{ \opp_n }{ \opr_p \opr_q } }_0
    -i \omega_m \expval{ \comm{ \opp_m }{ \opr_p \opr_q } \opr_n }_0 \nnnl
    =& - (\delta_{m,p}\delta_{n,q} + \delta_{m,q}\delta_{n,p})
    \left[ \omega_m \left( n_n + \frac{1}{2} \right) + \omega_n \left( n_m + \frac{1}{2} \right) \right]
\end{align}

\begin{align} \label{eq:s_der_2nd_comm_xx_pp}
    \expval{ \comm{ \comm{\opH}{\opr_m\opr_n} }{ \opp_p \opp_q } }_0
    =& \expval{ \comm{ -i (\omega_n \opr_m \opp_n + \omega_m \opp_m \opr_n) }{ \opp_p \opp_q } }_0 \nnnl
    =& -i \omega_n \expval{ \comm{ \opr_m }{ \opp_p \opp_q } \opp_n }_0
    -i \omega_m \expval{ \opp_m \comm{ \opr_n }{ \opp_p \opp_q } }_0 \nnnl
    =& (\delta_{m,p}\delta_{n,q} + \delta_{m,q}\delta_{n,p})
    \left[ \omega_m \left( n_m + \frac{1}{2} \right) + \omega_n \left( n_n + \frac{1}{2} \right) \right]
\end{align}

\begin{align} \label{eq:s_der_2nd_comm_pp_xx}
    \expval{ \comm{ \comm{\opH}{\opp_m\opp_n} }{ \opr_p \opr_q } }_0
    =& \expval{ \comm{ i \opp_m \dfdx{\opV}{r_n} + i \dfdx{\opV}{r_m} \opp_n }{ \opr_p \opr_q } }_0 \nnnl
    =& i \expval{ \comm{ \opp_m }{ \opr_p \opr_q } \dfdx{\opV}{r_n} }_0
    + i \expval{ \dfdx{\opV}{r_m} \comm{ \opp_n }{ \opr_p \opr_q } }_0 \nnnl
    =& \delta_{m,p} \expval{ \opr_q \dfdx{\opV}{r_n} }_0
    + \delta_{m,q} \expval{ \opr_p \dfdx{\opV}{r_n} }_0
    + \delta_{n,p} \expval{ \dfdx{\opV}{r_m} \opr_q }_0
    + \delta_{n,q} \expval{ \dfdx{\opV}{r_m} \opr_p }_0 \nnnl
    =& (\delta_{m,p}\delta_{n,q} + \delta_{m,q}\delta_{n,p})
    \left[ \omega_m \left( n_m + \frac{1}{2} \right) + \omega_n \left( n_n + \frac{1}{2} \right) \right]
\end{align}

\begin{align} \label{eq:s_der_2nd_comm_pp_pp}
    \expval{ \comm{ \comm{\opH}{\opp_m\opp_n} }{ \opp_p \opp_q } }_0
    =& \expval{ \comm{ i \opp_m \dfdx{\opV}{r_n} + i \dfdx{\opV}{r_m} \opp_n }{ \opp_p \opp_q } }_0 \nnnl
    =& i \expval{ \opp_m \comm{ \dfdx{\opV}{r_n} }{ \opp_p \opp_q } }_0
    + i \expval{ \comm{ \dfdx{\opV}{r_m} }{ \opp_p \opp_q } \opp_n }_0 \nnnl
    =& - \expval{ \opp_m \opp_p \dfdxy{\opV}{r_n}{r_q} }_0
    - \expval{ \opp_m \dfdxy{\opV}{r_n}{r_p} \opp_q }_0
    - \expval{ \opp_p \dfdxy{\opV}{r_m}{r_q} \opp_n }_0
    - \expval{ \dfdxy{\opV}{r_m}{r_p} \opp_q \opp_n }_0 \nnnl
    =& - \left[ \left(
    \expval{ \opp_m \dfdxy{\opV}{r_n}{r_q} \opp_p }_0
    + (m \leftrightarrow n) \right) + (p \leftrightarrow q) \right]
    + i \expval{ \opp_m \dfdxyz{\opV}{r_n}{r_q}{r_p} }_0
    - i \expval{ \dfdxyz{\opV}{r_m}{r_p}{r_q} \opp_n }_0 \nnnl
    =& - \left[ \left( \frac{1}{4}\Phi^{(4)}_{mnpq} + \delta_{m,p} \delta_{n,q} \omega_n \left( n_m + \frac{1}{2} \right)
    + (m \leftrightarrow n) \right) + (p \leftrightarrow q) \right]
    + \Phi^{(4)}_{mnpq} \nnnl
    =& - (\delta_{m,p}\delta_{n,q} + \delta_{m,q}\delta_{n,p})
    \left[ \omega_m \left( n_n + \frac{1}{2} \right) + \omega_n \left( n_m + \frac{1}{2} \right) \right]
\end{align}
In the fifth equality of \myeqref{eq:s_der_2nd_comm_pp_pp}, we used \myeqref{eq:s_der_iden_ppO}.

\begin{align} \label{eq:s_der_2nd_comm_xp_xp}
    \expval{ \comm{ \comm{\opH}{\opr_m\opp_n} }{ \opr_p \opp_q } }_0
    =& \expval{ \comm{ i \opr_m \dfdx{\opV}{r_n} - i\omega_m \opp_m \opp_n }{ \opr_p \opp_q } }_0 \nnnl
    =& i \expval{ \opr_p \comm{ \opr_m \dfdx{\opV}{r_n} }{ \opp_q } }_0 
    -i \omega_m \expval{ \comm{ \opp_m \opp_n }{ \opr_p } \opp_q }_0 \nnnl
    =& - \expval{ \opr_p \left( \delta_{m,q} \dfdx{\opV}{r_n} + \opr_m \dfdxy{\opV}{r_n}{r_q} \right) }_0
    - \omega_m \left[ \delta_{m,p} \delta_{n,q} \left( n_n + \frac{1}{2} \right) + \delta_{m,q} \delta_{n,p} \left( n_m + \frac{1}{2} \right) \right] \nnnl
    =& - \delta_{m,q}\delta_{n,p} \left[ \omega_m  \left( n_m + \frac{1}{2} \right) + \omega_n \left( n_n + \frac{1}{2} \right) \right]
    - \delta_{m,p}\delta_{n,q} \left[ \omega_m \left( n_n + \frac{1}{2} \right) + \omega_n\left( n_m + \frac{1}{2} \right) \right] \nnnl
    &- \left( n_p + \frac{1}{2} \right)\left( n_m + \frac{1}{2} \right) \Phi^{(4)}_{mnpq},
\end{align}

\begin{align} \label{eq:s_der_2nd_comm_xp_xx}
    \expval{ \comm{ \comm{\opH}{\opr_m\opp_n} }{ \opr_p \opr_q } }_0
    =& \expval{ \comm{ i \opr_m \dfdx{\opV}{r_n} - i\omega_m \opp_m \opp_n }{ \opr_p \opr_q } }_0 \nnnl
    =& -i \omega_m \left( \expval{ \comm{ \opp_m \opp_n }{ \opr_p } \opr_q }_0 + \expval{ \opr_p \comm{ \opp_m \opp_n }{ \opr_q } }_0 \right) \nnnl
    =& 0
\end{align}

\begin{align} \label{eq:s_der_2nd_comm_xp_pp}
    \expval{ \comm{ \comm{\opH}{\opr_m\opp_n} }{ \opp_p \opp_q } }_0
    =& \expval{ \comm{ i \opr_m \dfdx{\opV}{r_n} - i\omega_m \opp_m \opp_n }{ \opp_p \opp_q } }_0 \nnnl
    =& i \expval{ \opp_p \comm{ \opr_m \dfdx{\opV}{r_n} }{ \opp_q } }_0 + i \expval{ \comm{ \opr_m \dfdx{\opV}{r_n} }{ \opp_p } \opp_q }_0 \nnnl
    =& - \expval{ \opp_p \left ( \delta_{m,q} \dfdx{\opV}{r_n} + \opr_m \dfdxy{\opV}{r_n}{r_q} \right) }_0 - \expval{ \left ( \delta_{m,p} \dfdx{\opV}{r_n} + \opr_m \dfdxy{\opV}{r_n}{r_p} \right) \opp_q }_0 \nnnl
    =& 0
\end{align}

\begin{align} \label{eq:s_der_2nd_comm_xx_xp}
    \expval{ \comm{ \comm{\opH}{\opr_m\opr_n} }{ \opr_p \opp_q } }_0
    =& -i \expval{ \comm{ (\omega_n \opr_m \opp_n + \omega_m \opp_m \opr_n)}{ \opr_p \opp_q } }_0 \nnnl
    =& -i \omega_n \expval{ \comm{ \opr_m \opp_n}{ \opr_p \opp_q } }_0
    -i \omega_m \expval{ \comm{ \opp_m \opr_n}{ \opr_p \opp_q } }_0 \nnnl
    =& -i \omega_n \expval{ -i \delta_{n,p} \opr_m \opp_q + i \delta_{m,q} \opr_p \opp_n }_0
    -i \omega_m \expval{ i \delta_{n,q} \opp_m \opr_p -i \delta_{m,p} \opp_q \opr_n }_0 \nnnl
    =& \frac{i}{2} \omega_n( - \delta_{n,p} \delta_{m,q} + \delta_{m,q} \delta_{p,n})
    + \frac{i}{2} \omega_m (- \delta_{n,q} \delta_{m,p} + \delta_{m,p} \delta_{q,n}) \nnnl
    =& 0
\end{align}

\begin{align} \label{eq:s_der_2nd_comm_pp_xp}
    \expval{ \comm{ \comm{\opH}{\opp_m\opp_n} }{ \opr_p \opp_q } }_0
    =& \expval{ \comm{ i \opp_m \dfdx{\opV}{r_n} + i \dfdx{\opV}{r_m} \opp_n }{ \opr_p \opp_q } }_0 \nnnl
    =& i\expval{ \opp_m \opr_p \comm{ \dfdx{\opV}{r_n} }{ \opp_q } }_0
    + i\expval{ \comm{ \opp_m  }{ \opr_p } \opp_q \dfdx{\opV}{r_n} }_0
    + i \expval{ \dfdx{\opV}{r_m} \comm{ \opp_n }{ \opr_p } \opp_q }_0
    + i \expval{ \opr_p \comm{ \dfdx{\opV}{r_m} }{ \opp_q } \opp_n }_0 \nnnl
    =& -\expval{ \opp_m \opr_p \dfdxy{\opV}{r_n}{r_q} }_0
    + \delta_{m,p} \expval{\opp_q \dfdx{\opV}{r_n} }_0
    + \delta_{n,p} \expval{ \dfdx{\opV}{r_m} \opp_q }_0
    - \expval{ \opr_p \dfdxy{\opV}{r_m}{r_q} \opp_n }_0 \nnnl
    =& \frac{i}{2} \expval{\delta_{p,m} \dfdxy{\opV}{r_n}{r_q}
    + \opr_p \dfdxyz{\opV}{r_n}{r_q}{r_m} }_0
    - \frac{i}{2} \delta_{m,p} \expval{\dfdxy{\opV}{r_n}{r_q} }_0 \nnnl
    +& \frac{i}{2} \delta_{n,p} \expval{ \dfdxy{\opV}{r_m}{r_q} }_0
    - \frac{i}{2} \expval{ \delta_{p,n} \dfdxy{\opV}{r_m}{r_q}
    + \opr_p \dfdxyz{\opV}{r_m}{r_q}{r_n} }_0 \nnnl
    =& 0
\end{align}

Now, we actually calculate the second derivatives of energy.
Using \myeqref{eq:s_der_ddE_1}, and Eqs.~(\ref{eq:s_der_2nd_comm_xx_xx}-\ref{eq:s_der_2nd_comm_pp_pp}), one finds
\begin{align}
    \dfdxy{E}{\beti{mn}}{\beti{pq}}
    =& -\frac{1}{2} b_{mn} b_{pq} \expval{ \comm{ \comm{\opH}{\opr_m\opr_n - \opp_m\opp_n} }{\opr_p\opr_q - \opp_p\opp_q } }_0
    + ((m,n) \leftrightarrow (p,q)) \nnnl
    =& 2 b_{mn} b_{pq} (\delta_{m,p}\delta_{n,q} + \delta_{m,q}\delta_{n,p})
    \left[ \omega_m (n_n + n_m + 1) + \omega_n (n_m + n_n + 1) \right] \nnnl
    =& (\omega_m + \omega_n) \delta_{mn,pq},
\end{align}
where in the last equality we have used $m\leq n$ and $p\leq q$,
\begin{align} \label{eq:s_der_2nd_ci_ci}
    \dfdxy{E}{\gami{mn}}{\gami{pq}}
    =& -\frac{1}{2} c_{mn} c_{pq} \expval{ \comm{ \comm{\opH}{\opr_m\opr_n + \opp_m\opp_n} }{\opr_p\opr_q + \opp_p\opp_q } }_0
    + ((m,n) \leftrightarrow (p,q)) \nnnl
    =& 2 c_{mn} c_{pq} (\delta_{m,p}\delta_{n,q} + \delta_{m,q}\delta_{n,p})
    \left[ \omega_m (n_n - n_m) + \omega_n (n_m - n_n) \right] \nnnl
    =& ( \omega_n - \omega_m) \delta_{mn,pq},
\end{align}
where in the last equality we have used $m<n$ and $p<q$, and
\begin{align}
    \dfdxy{E}{\beti{mn}}{\gami{pq}}
    = \frac{1}{2} b_{mn} c_{pq} \left( \expval{ \comm{ \comm{\opH}{\opr_m\opr_n - \opp_m\opp_n} }{\opr_p\opr_q + \opp_p\opp_q } }_0
    + \expval{ \comm{ \comm{\opH}{\opr_p\opr_q + \opp_p\opp_q} }{\opr_m\opr_n - \opp_m\opp_n} }_0 \right)
    = 0.
\end{align}

Also, using \myeqref{eq:s_der_2nd_comm_xp_xp}, one finds
\begin{align}
    \dfdxy{E}{\betr{mn}}{\betr{pq}}
    =& -\frac{1}{2} b_{mn} b_{pq} \expval{ \comm{ \comm{\opH}{\opr_m\opp_n + \opp_m\opr_n} }{\opr_p\opp_q + \opp_p\opr_q } }_0
    + ((m,n) \leftrightarrow (p,q)) \nnnl
    =& - b_{mn}b_{pq}\left[ \left( \expval{ \comm{ \comm{\opH}{\opr_m\opp_n} }{ \opr_p \opp_q } }_0 + (p \leftrightarrow q) \right) + (m \leftrightarrow n)\right] \nnnl
    =& - b_{mn}b_{pq} \Big[ - (\delta_{m,p}\delta_{n,q} + \delta_{m,q}\delta_{n,p}) (\omega_m + \omega_n) (n_m + n_n + 1)
    - (n_p + n_q + 1)\left( n_m + \frac{1}{2} \right) \Phi^{(4)}_{mnpq}  + (m \leftrightarrow n) \Big] \nnnl
    =& 2 b_{mn}b_{pq} (\delta_{m,p}\delta_{n,q} + \delta_{m,q}\delta_{n,p}) (\omega_m + \omega_n) (n_m + n_n + 1)
    + b_{mn}b_{pq}(n_p + n_q + 1)(n_m + n_n + 1) \Phi^{(4)}_{mnpq} \nnnl
    =& (\omega_m + \omega_n) \delta_{mn,pq} + b_{mn}b_{pq}(n_p + n_q + 1)(n_m + n_n + 1) \Phi^{(4)}_{mnpq},
\end{align}
\begin{align}
    \dfdxy{E}{\betr{mn}}{\gamr{pq}}
    =& \frac{1}{2} b_{mn} c_{pq} \left(\expval{ \comm{ \comm{\opH}{\opr_m\opp_n + \opp_m\opr_n} }{\opr_p\opp_q - \opp_p\opr_q } }_0
    + \expval{ \comm{ \comm{\opH}{\opr_p\opp_q - \opp_p\opr_q} }{\opr_m\opp_n + \opp_m\opr_n} }_0 \right) \nnnl
    =& b_{mn}c_{pq}\left[ \left( \expval{ \comm{ \comm{\opH}{\opr_m\opp_n} }{ \opr_p \opp_q } }_0 - (p \leftrightarrow q) \right) + (m \leftrightarrow n)\right] \nnnl
    =& b_{mn}c_{pq} \Big[ (\delta_{m,p}\delta_{n,q} - \delta_{m,q}\delta_{n,p}) (\omega_m - \omega_n) (n_m - n_n)
    - (n_p - n_q)\left( n_m + \frac{1}{2} \right) \Phi^{(4)}_{mnpq} + (m \leftrightarrow n) \Big] \nnnl
    =& - b_{mn}c_{pq} (n_m + n_n + 1)(n_p - n_q) \Phi^{(4)}_{mnpq},
\end{align}
and
\begin{align} \label{eq:s_der_2nd_cr_cr}
    \dfdxy{E}{\gamr{mn}}{\gamr{pq}}
    =& -\frac{1}{2} c_{mn} c_{pq} \expval{ \comm{ \comm{\opH}{\opr_m\opp_n - \opp_m\opr_n} }{\opr_p\opp_q - \opp_p\opr_q } }_0
    + ((m,n) \leftrightarrow (p,q)) \nnnl
    =& - c_{mn}c_{pq}\left[ \left( \expval{ \comm{ \comm{\opH}{\opr_m\opp_n} }{ \opr_p \opp_q } }_0 - (p \leftrightarrow q) \right) - (m \leftrightarrow n)\right] \nnnl
    =& - c_{mn}c_{pq} \Big[ (\delta_{m,p}\delta_{n,q} - \delta_{m,q}\delta_{n,p}) (\omega_m - \omega_n) (n_m - n_n)
    - (n_p - n_q)\left( n_m + \frac{1}{2} \right) \Phi^{(4)}_{mnpq} - (m \leftrightarrow n) \Big] \nnnl
    =& -2 c_{mn}c_{pq} \delta_{mn,pq} (\omega_m - \omega_n) (n_m - n_n) 
    + c_{mn}c_{pq} (n_p - n_q) (n_m - n_n) \Phi^{(4)}_{mnpq} \nnnl
    =& (\omega_n - \omega_m) \delta_{mn,pq}
    + c_{mn}c_{pq} (n_p - n_q) (n_m - n_n) \Phi^{(4)}_{mnpq}.
\end{align}

Using Eqs.~(\ref{eq:s_der_2nd_comm_xp_xx}-\ref{eq:s_der_2nd_comm_pp_xp}), one finds
\begin{align}
    \dfdxy{E}{\betr{mn}}{\beti{pq}}
    = \frac{1}{2} b_{mn} b_{pq} \left(\expval{ \comm{ \comm{\opH}{\opr_m\opp_n + \opp_m\opr_n} }{\opr_p\opr_q - \opp_p\opp_q } }_0
    + \expval{ \comm{ \comm{\opH}{\opr_p\opr_q - \opp_p\opp_q } }{\opr_m\opp_n + \opp_m\opr_n} }_0 \right)
    = 0,
\end{align}
\begin{align}
    \dfdxy{E}{\betr{mn}}{\gami{pq}}
    = - \frac{1}{2} b_{mn} c_{pq} \left(\expval{ \comm{ \comm{\opH}{\opr_m\opp_n + \opp_m\opr_n} }{\opr_p\opr_q + \opp_p\opp_q } }_0
    + \expval{ \comm{ \comm{\opH}{\opr_p\opr_q + \opp_p\opp_q } }{\opr_m\opp_n + \opp_m\opr_n} }_0 \right)
    = 0,
\end{align}
\begin{align}
    \dfdxy{E}{\gamr{mn}}{\beti{pq}}
    = -\frac{1}{2} c_{mn} b_{pq} \left(\expval{ \comm{ \comm{\opH}{\opr_m\opp_n - \opp_m\opr_n} }{\opr_p\opr_q - \opp_p\opp_q } }_0
    + \expval{ \comm{ \comm{\opH}{\opr_p\opr_q - \opp_p\opp_q } }{\opr_m\opp_n - \opp_m\opr_n} }_0 \right)
    = 0,
\end{align}
and
\begin{align}
    \dfdxy{E}{\gamr{mn}}{\gami{pq}}
    = \frac{1}{2} c_{mn} c_{pq} \left(\expval{ \comm{ \comm{\opH}{\opr_m\opp_n - \opp_m\opr_n} }{\opr_p\opr_q + \opp_p\opp_q } }_0
    + \expval{ \comm{ \comm{\opH}{\opr_p\opr_q + \opp_p\opp_q } }{\opr_m\opp_n - \opp_m\opr_n} }_0 \right)
    = 0.
\end{align}

Finally, for mixed second derivatives for one displacement and one squeezing parameter, the relevant expectation values are
\begin{align}
    \expval{ \comm{ \comm{\opH}{\opr_p} }{ \opr_m\opr_n } }_0
    = -i \omega_p \expval{ \comm{ \opp_p }{ \opr_m\opr_n } }_0
    = 0,
\end{align}
\begin{align}
    \expval{ \comm{ \comm{\opH}{\opr_p} }{ \opr_m\opp_n } }_0
    = -i \omega_p \expval{ \comm{ \opp_p }{ \opr_m\opp_n } }_0
    = 0,
\end{align}
\begin{align}
    \expval{ \comm{ \comm{\opH}{\opr_p} }{ \opp_m\opp_n } }_0
    = -i \omega_p \expval{ \comm{ \opp_p }{ \opp_m\opp_n } }_0
    = 0,
\end{align}
\begin{align}
    \expval{ \comm{ \comm{\opH}{\opp_p} }{ \opr_m\opr_n } }_0
    = i \expval{ \comm{ \dfdx{\opV}{r_p} }{ \opr_m\opr_n } }_0
    = 0,
\end{align}
\begin{align}
    \expval{ \comm{ \comm{\opH}{\opp_p} }{ \opr_m\opp_n } }_0
    = i \expval{ \comm{ \dfdx{\opV}{r_p} }{ \opr_m\opp_n } }_0 
    = i \expval{ \opr_m \comm{ \dfdx{\opV}{r_p} }{ \opp_n } }_0 
    = - \expval{ \opr_m \dfdxy{\opV}{r_p}{r_n} }_0 
    = - \left( n_m + \frac{1}{2} \right) \Phi^{(3)}_{mnp},
\end{align}
and
\begin{align}
    \expval{ \comm{ \comm{\opH}{\opp_p} }{ \opp_m\opp_n } }_0
    =& i \expval{ \comm{ \dfdx{\opV}{r_p} }{ \opp_m\opp_n } }_0 \nnnl
    =& i \expval{ \opp_m \comm{ \dfdx{\opV}{r_p} }{ \opp_n } }_0 + i \expval{ \comm{ \dfdx{\opV}{r_p} }{ \opp_m } \opp_n }_0 \nnnl
    =& - \expval{ \opp_m \dfdxy{\opV}{r_p}{r_n} }_0 - \expval{ \dfdxy{\opV}{r_p}{r_m} \opp_n }_0 \nnnl
    =& \frac{i}{2}\Phi^{(3)}_{mnp} -\frac{i}{2} \Phi^{(3)}_{mnp} \nnnl
    =& 0.
\end{align}

Using \myeqref{eq:s_der_ddE_2}, the mixed second derivatives of the energy become
\begin{align}
    \dfdxy{E}{\alpr{p}}{\betr{mn}}
    = - b_{mn} \expval{ \comm{ \comm{\opH}{\opp_p} }{ \opr_m\opp_n + \opp_m\opr_n } }_0
    = b_{mn} \left[ \left( n_m + \frac{1}{2} \right) \Phi^{(3)}_{mnp} + (n \leftrightarrow m) \right]
    =& b_{mn} (n_m + n_n + 1) \Phi^{(3)}_{mnp},
\end{align}
\begin{align}
    \dfdxy{E}{\alpr{p}}{\gamr{mn}}
    = c_{mn} \expval{ \comm{ \comm{\opH}{\opp_p} }{ \opr_m\opp_n - \opp_m\opr_n } }_0
    = - c_{mn} \left[ \left( n_m + \frac{1}{2} \right) \Phi^{(3)}_{mnp} - (n \leftrightarrow m) \right]
    =& - c_{mn} (n_m - n_n) \Phi^{(3)}_{mnp},
\end{align}
\begin{align}
    \dfdxy{E}{\alpr{p}}{\beti{mn}}
    = b_{mn} \expval{ \comm{ \comm{\opH}{\opp_p} }{ \opr_m\opr_n - \opp_m\opp_n } }_0
    = 0,
\end{align}
\begin{align}
    \dfdxy{E}{\alpr{p}}{\gami{mn}}
    = -c_{mn} \expval{ \comm{ \comm{\opH}{\opp_p} }{ \opr_m\opr_n + \opp_m\opp_n } }_0
    = 0,
\end{align}
\begin{align}
    \dfdxy{E}{\alpi{p}}{\betr{mn}}
    = b_{mn} \expval{ \comm{ \comm{\opH}{\opr_p} }{ \opr_m\opp_n + \opp_m\opr_n } }_0
    = 0
\end{align}
\begin{align}
    \dfdxy{E}{\alpi{p}}{\gamr{mn}}
    = -c_{mn} \expval{ \comm{ \comm{\opH}{\opr_p} }{ \opr_m\opp_n - \opp_m\opr_n } }_0
    = 0,
\end{align}
\begin{align}
    \dfdxy{E}{\alpi{p}}{\beti{mn}}
    = -b_{mn} \expval{ \comm{ \comm{\opH}{\opr_p} }{ \opr_m\opr_n - \opp_m\opp_n } }_0
    = 0,
\end{align}
and
\begin{align}
    \dfdxy{E}{\alpi{p}}{\gami{mn}}
    = c_{mn} \expval{ \comm{ \comm{\opH}{\opr_p} }{ \opr_m\opr_n + \opp_m\opp_n } }_0
    = 0.
\end{align}

\section{Calculation of the interacting Green function} \label{sec:supp_green}
In this section, we detail the solution of the Dyson equations.

\subsection{Partially interacting Green function: 4-phonon interaction}
First, let us consider the Dyson equation for the partially interacting Green function [\eqref{eq:anh_G4_2_dyson}].
Substituting \myeqref{eq:anh_V4_def} and \myeqref{eq:anh_G0} into \myeqref{eq:anh_G4_2_dyson}, one can directly solve the Dyson equation to find
\begin{align} \label{eq:anh_G4_solution}
    \mbPII \mbmcG[4](z) \mbPII 
    = 0 \oplus (\mbmcG[0]_{2+} \oplus \mbmcG[0]_{2-})
    \times \begin{pmatrix}
			\one & 0 & 0 & 0\\
			iB \mbPhf B \frac{z}{z^2 - \mbom_{+}^2}
			& \one - B \mbPhf B \frac{\mbom_{+}}{z^2 - \mbom_{+}^2}
			& iB \mbPhf C \frac{z}{z^2 - \mbom_{-}^2}
			& -B \mbPhf C \frac{\mbom_{-}}{z^2 - \mbom_{-}^2} \\
			0 & 0 & \one & 0 \\
			iC \mbPhf B \frac{z}{z^2 - \mbom_{+}^2}
			& -C \mbPhf B \frac{\mbom_{+}}{z^2 - \mbom_{+}^2}
			& iC \mbPhf C \frac{z}{z^2 - \mbom_{-}^2}
			& \one - C \mbPhf C \frac{\mbom_{-}}{z^2 - \mbom_{-}^2}
	\end{pmatrix}^{-1}.
\end{align}
From \myeqref{eq:anh_G4_solution}, one finds
\begin{widetext}
\begin{align} \label{eq:s_green_G4_equation}
    \begin{pmatrix}
        \mbmcG[0]_{2+}(z) & 0 \\
        0 & \mbmcG[0]_{2-}(z)
    \end{pmatrix}
    =
    \begin{pmatrix}
        \mbmcG[4]_{++}(z) & \mbmcG[4]_{+-}(z) \\
        \mbmcG[4]_{+-}(z) & \mbmcG[4]_{--}(z)
    \end{pmatrix}
    \times
    \begin{pmatrix}
		\one & 0 & 0 & 0\\
		iB \mbPhf B \frac{z}{z^2 - \mbom_{+}^2}
		& \one - B \mbPhf B \frac{\mbom_{+}}{z^2 - \mbom_{+}^2}
		& iB \mbPhf C \frac{z}{z^2 - \mbom_{-}^2}
		& -B \mbPhf C \frac{\mbom_{-}}{z^2 - \mbom_{-}^2} \\
		0 & 0 & \one & 0 \\
		iC \mbPhf B \frac{z}{z^2 - \mbom_{+}^2}
		& -C \mbPhf B \frac{\mbom_{+}}{z^2 - \mbom_{+}^2}
		& iC \mbPhf C \frac{z}{z^2 - \mbom_{-}^2}
		& \one - C \mbPhf C \frac{\mbom_{-}}{z^2 - \mbom_{-}^2}
	\end{pmatrix},
\end{align}
where we defined
\begin{equation}
    \mbmcG[4]_{ss'}(z) = \mathbf{P}_{2s} \mbmcG[4](z) \mathbf{P}_{2s'}
\end{equation}
with $s,s' \in \{ +, - \}$.

By explicitly writing the odd rows and even columns of \myeqref{eq:s_green_G4_equation}, one finds
\begin{align} \label{eq:s_green_G4_12}
    \begin{pmatrix}
        \frac{i \mbom_{+}}{z^2 - \mbom_{+}^2} & 0 \\
        0 & \frac{i \mbom_{-}}{z^2 - \mbom_{-}^2}
    \end{pmatrix}
    =
    \begin{pmatrix}
        [\mbmcG[4]_{++}(z)]_{11} & [\mbmcG[4]_{++}(z)]_{12} &
        [\mbmcG[4]_{+-}(z)]_{11} & [\mbmcG[4]_{+-}(z)]_{12} \\
        [\mbmcG[4]_{-+}(z)]_{11} & [\mbmcG[4]_{-+}(z)]_{12} &
        [\mbmcG[4]_{--}(z)]_{11} & [\mbmcG[4]_{--}(z)]_{12}
    \end{pmatrix}
    \times 
    \begin{pmatrix}
		0 & 0 \\
		\one - B \mbPhf B \frac{\mbom_{+}}{z^2 - \mbom_{+}^2}
		& -B \mbPhf C \frac{\mbom_{-}}{z^2 - \mbom_{-}^2} \\
		0 & 0 \\
		-C \mbPhf B \frac{\mbom_{+}}{z^2 - \mbom_{+}^2}
		& \one - C \mbPhf C \frac{\mbom_{-}}{z^2 - \mbom_{-}^2}
	\end{pmatrix}.
\end{align}
Here, the subscript $11$ and $12$ denotes the row and column index of the blocks in the $2\times 2$ representation of $\mbmcG[4]_{ss'}(z)$.
Since the first and third rows of the last matrix of \myeqref{eq:s_green_G4_12} is zero, one finds
\begin{align} \label{eq:s_green_G4_12_modified}
    \begin{pmatrix}
        \frac{i \mbom_{+}}{z^2 - \mbom_{+}^2} & 0 \\
        0 & \frac{i \mbom_{-}}{z^2 - \mbom_{-}^2}
    \end{pmatrix}
    =
    \begin{pmatrix}
        [\mbmcG[4]_{++}(z)]_{12} & [\mbmcG[4]_{+-}(z)]_{12} \\
        [\mbmcG[4]_{-+}(z)]_{12} & [\mbmcG[4]_{--}(z)]_{12}
    \end{pmatrix}
    \times 
    \begin{pmatrix}
		\one - B \mbPhf B \frac{\mbom_{+}}{z^2 - \mbom_{+}^2}
		& -B \mbPhf C \frac{\mbom_{-}}{z^2 - \mbom_{-}^2} \\
		-C \mbPhf B \frac{\mbom_{+}}{z^2 - \mbom_{+}^2}
		& \one - C \mbPhf C \frac{\mbom_{-}}{z^2 - \mbom_{-}^2}
	\end{pmatrix}.
\end{align}
\end{widetext}
By inverting the last matrix of \myeqref{eq:s_green_G4_12_modified} and using \myeqref{eq:anh_g_def}, one finds
\begin{align} \label{eq:anh_G4_2_xx}
\begin{pmatrix}
[\mbmcG[4]_{++}(z)]_{12} & [\mbmcG[4]_{+-}(z)]_{12} \\
[\mbmcG[4]_{-+}(z)]_{12} & [\mbmcG[4]_{--}(z)]_{12}
\end{pmatrix}
=& i
\begin{pmatrix}
\mb{g}_+(z) & 0\\
0 & \mb{g}_-(z)
\end{pmatrix}
\left[ \one -
\begin{pmatrix}
  B \mbPhf  B \mb{g}_+(z)
& B \mbPhf  C \mb{g}_-(z)\\
  C \mbPhf  B \mb{g}_+(z)
& C \mbPhf  C \mb{g}_-(z)
\end{pmatrix} \right]^{-1}
\end{align}
where
\begin{equation} \label{eq:anh_g_def}
    \mb{g}_\pm(z) = \frac{\mbom_\pm}{z^2 - \mbom_\pm^2}.
\end{equation}

\subsection{Fully interacting Green function: 3-, 4-phonon interactions} \label{sec:supp_green_fully}
Next, we derive the Dyson equation for the interacting retarded position-position correlation function starting from the Dyson equation in \myeqref{eq:anh_G1_Dyson}.

Using \myeqref{eq:s_transf_op_der} and Eqs.~(\ref{eq:s_overlap_dU_ar}-\ref{eq:s_overlap_dU_ci}), one can easily show that the matrix elements of the position operator is nonzero only for the variation of $\alpr{m}$:
\begin{equation}
    \partial_\mu \mb{r} = \begin{pmatrix} \one & 0 & 0 & 0 & 0 & 0 \end{pmatrix}^\intercal.
\end{equation}
Similarly, the matrix element for the momentum operator is nonzero only for the variation of $\alpi{m}$:
\begin{equation}
    \partial_\mu \mb{p} = \begin{pmatrix} 0 & \one & 0 & 0 & 0 & 0 \end{pmatrix}^\intercal,
\end{equation}
By substituting $\opr_m$ or $\opp_n$ to $\hat{A}$ and $\hat{B}$ of the general linear response formula \myeqref{eq:tdvp_lr_eqn}, we find
\begin{equation} \label{eq:s_green_corr}
    \mbPI \mbmcG(\omega + i\eta) \mbPI
    = i \begin{pmatrix}
    -\mb{G}^{\rm (R)}_{rp}(\omega) & \mb{G}^{\rm (R)}_{rr}(\omega) \\
    -\mb{G}^{\rm (R)}_{pp}(\omega) & \mb{G}^{\rm (R)}_{pr}(\omega)
    \end{pmatrix}
    \oplus \mathbb{0}.
\end{equation}
Therefore, to calculate the position-position correlation function, it suffices to compute the upper right block of $\mbPI \mbmcG \mbPI$, the 1-phonon sector of the fully interacting Green function.

By viewing the upper right block of \myeqref{eq:anh_G1_Dyson}, one finds
\begin{align} \label{eq:s_green_G3_rr}
    i \mb{G}^\mathrm{(R)}_{rr}
    = i \mb{G}^\mathrm{(R0)}_{rr}
    - i \mb{G}^\mathrm{(R0)}_{rr}
    \times \left( \sum_{s,s'=\pm} \mbPht B_s [\mbmcG[4]_{ss'}(z)]_{12} B_{s'} \mbPht \right)
    \times i \mb{G}^\mathrm{(R)}_{rr}.
\end{align}
Then, the self-energy for the fully interacting retarded position-position correlation function reads
\begin{equation} \label{eq:s_green_Pi_def}
    \mb{\Pi}_{rr}(z) = -i \sum_{s,s'=\pm} \mbPht B_s [\mbmcG[4]_{ss'}(z)]_{12} B_{s'} \mbPht.
\end{equation}

Substituting \myeqref{eq:anh_G4_2_xx} into \myeqref{eq:s_green_Pi_def} and using
\begin{align}
    \begin{pmatrix} B \mb{g}_+ & C \mb{g}_- \end{pmatrix}
    \left[ \one -
    \begin{pmatrix} B \\ C \end{pmatrix}
    \mbPhf
    \begin{pmatrix} B \mb{g}_+ & C \mb{g}_- \end{pmatrix}
    \right]^{-1}
    \begin{pmatrix} B \\ C \end{pmatrix}
    = (B \mb{g}_+ B + C \mb{g}_- C)
    \left[ \one - \mbPhf (B \mb{g}_+ B + C \mb{g}_- C) \right]^{-1},
\end{align}
we find \myeqref{eq:anh_Pi} of the main text:
\begin{align} \label{eq:s_green_anh_Pi}
\mb{\Pi}_{rr}(z)
=& \mbPht \mb{W} (\one - \mbPhf \mb{W})^{-1} \mbPht.
\end{align}
Here, we used the diagonal matrix $\mb{W}$ defined in \myeqref{eq:anh_W_def} of the main text:
\begin{align} \label{eq:s_green_W_def}
W_{mn,pq}
&\equiv \left[ B \mb{g}_+(z) B + C \mb{g}_-(z) C \right]_{mn, pq} \nnnl
&= -\frac{1}{2} \, \frac{2 - \delta_{m,n}}{2} \Bigg[
 \frac{(\omega_m + \omega_n)(n_m + n_n + 1)}{(\omega_m + \omega_n)^2 - z^2}
- \frac{(\omega_m - \omega_n)(n_m - n_n)}{(\omega_m - \omega_n)^2 - z^2}
\Bigg] \delta_{mn, pq}.
\end{align}
In \myeqref{eq:anh_Pi}, all the sum over indices in the matrix-matrix product should be constrained by $m \leq n$.

\section{Derivation of the SCHA {\ansatz} from the TDVP self-energy} \label{sec:supp_selfen}
In this section, we derive the SCHA {\ansatz} \myeqref{eq:anh_Pi_chi} from the self-energy formula \myeqref{eq:anh_Pi} which is derived from TDVP.
In this section, the constraint $m \leq n$ in the summation over mode indices $m$ and $n$ is not implied. The constraint is made explicit whenever necessary by using smaller matrices which are defined only on the constrained indices:
\begin{equation}
    \widetilde{\Phi}^{(3)}_{p,m'n'} = \Phi^{(3)}_{pm'n'},
\end{equation}
\begin{equation}
    \widetilde{\Phi}^{(4)}_{m'n',r's'} = \Phi^{(4)}_{m'n'r's'},
\end{equation}
and
\begin{equation}
    \widetilde{W}_{m'n',r's'} = W_{m'n',r's'}.
\end{equation}
Here and in the remaining part of this section, we denote the constrained indices with primes: the index $m'n'$ implies the constraint $m' \leq n'$.
Using these smaller matrices, \myeqref{eq:anh_Pi} can be written as
\begin{align} \label{eq:supp_selfen_anh_Pi}
\mb{\Pi}_{rr}(z)
= \widetilde{\mb{\Phi}}^{(3)} \widetilde{\mb{W}} (\one - \widetilde{\mb{\Phi}}^{(4)} \widetilde{\mb{W}})^{-1} \widetilde{\mb{\Phi}}^{(3) \intercal}.
\end{align}

Next, we define a rectangular matrix $\mb{R}$ with matrix elements
\begin{equation}
    R_{m'n',rs} =
    \begin{cases}
    1 & \text{if $(r,s)=(m',n')$ or $(r,s)=(n',m')$} \\
    0 & \text{otherwise}
    \end{cases}.
\end{equation}
By multiplying $\mb{R}$ to the smaller matrices, one can recover the full matrix:
\begin{equation} \label{eq:supp_selfen_R_Phi3}
    \widetilde{\mb{\Phi}}^{(3)} \mb{R}
    = \mb{R}^\intercal \widetilde{\mb{\Phi}}^{(3) \intercal}
    = \mb{\Phi}^{(3)},
\end{equation}
and
\begin{equation} \label{eq:supp_selfen_R_Phi4}
    \mb{R}^\intercal \widetilde{\mb{\Phi}}^{(4)} \mb{R} = \mb{\Phi}^{(4)}.
\end{equation}
These identities hold because $\Phi^{(3)}_{pmn}$ and $\Phi^{(4)}_{mnrs}$ are invariant to the permutation of the indices.
In addition, from the definition of $\mb{\chi}$ [\myeqref{eq:anh_chi_def}], one finds
\begin{equation} \label{eq:supp_selfen_R_W_1}
    (\mb{R} \chi \mb{R}^\intercal)_{m'n',r's'}
    = (\mb{R} \chi \mb{R}^\intercal)_{m'n',m'n'} \delta_{m'n',r's'}
\end{equation}
and
\begin{align} \label{eq:supp_selfen_R_W_2}
    (\mb{R} \mb{\chi} \mb{R}^\intercal)_{m'n',m'n'}
    =& \begin{cases}
    \chi_{m'n'} & \text{if $m'=n'$} \\
    2\chi_{m'n'} & \text{if $m'\neq n'$}
    \end{cases} \nnnl
    =& (2 - \delta_{m,n}) \chi_{m'n'} \nnnl
    =& -2 \widetilde{W}_{m'n'}.
\end{align}
Equations \eqref{eq:supp_selfen_R_W_1} and \eqref{eq:supp_selfen_R_W_2} imply
\begin{equation} \label{eq:supp_selfen_R_W}
    \mb{R} \chi \mb{R}^\intercal
    = -2 \widetilde{\mb{W}}.
\end{equation}

Using Eqs.~\eqref{eq:supp_selfen_R_Phi3}, \eqref{eq:supp_selfen_R_Phi4}, and \eqref{eq:supp_selfen_R_W}, we can write \myeqref{eq:supp_selfen_anh_Pi} as
\begin{align} \label{eq:supp_selfen_anh_Pi_2}
\mb{\Pi}_{rr}(z)
=& -\frac{1}{2} \widetilde{\mb{\Phi}}^{(3)} \mb{R} \chi \mb{R}^\intercal \left( \one + \frac{1}{2} \widetilde{\mb{\Phi}}^{(4)} \mb{R} \chi \mb{R}^\intercal \right)^{-1} \widetilde{\mb{\Phi}}^{(3) \intercal} \nnnl
=& -\frac{1}{2} \widetilde{\mb{\Phi}}^{(3)} \mb{R} \chi \left( \one + \frac{1}{2} \mb{R}^\intercal \widetilde{\mb{\Phi}}^{(4)} \mb{R} \chi \right)^{-1} \mb{R}^\intercal \widetilde{\mb{\Phi}}^{(3) \intercal} \nnnl
=& -\frac{1}{2} \mb{\Phi}^{(3)} \chi \left( \one + \frac{1}{2} \mb{\Phi}^{(4)} \chi \right)^{-1} \mb{\Phi}^{(3)}.
\end{align}
Equation \eqref{eq:supp_selfen_anh_Pi_2} is identical to \myeqref{eq:anh_Pi_chi} of the main text.

\section{Degenerate and near-degenerate modes}
In this section, we detail the treatment of degenerate and near-degenerate modes.

First, let us assume that states $m$ and $n$ are almost but not exactly degenerate.
Although it is true that $c_{mn}$ [\myeqref{eq:s_tdvp_cmn_def}] becomes large, what is important in the dynamics of the variational parameter $\gamma_{mn}$ in the linear response regime is the linearized time-evolution generator $\mb{K}$.
When computing $\mb{K}$, the large $c_{mn}$ factors are counteracted by the small $(n_m - n_n)$ factors that appear when taking derivatives of $E(\mbx)$.
Equations~(\ref{eq:s_overlap_cr_ci}, \ref{eq:s_der_2nd_ci_ci}, \ref{eq:s_der_2nd_cr_cr}) are examples of such counteraction.
The fact that the equation of motion does not suffer any problems can be seen from Eqs.~(\ref{eq:anh_V3_def}, \ref{eq:anh_V4_def}, \ref{eq:anh_C_def}).
One finds that $c_{mn}$ enters the time-evolution generator $\mb{K}$ only through $C_{mn,pq}$ [\myeqref{eq:anh_C_def}].
There, the $c_{mn}$ factor is multiplied by $(n_m - n_n)$, so that $C_{mn,pq}$ is proportional to $\sqrt{n_m-n_n}$.
Hence, the matrix element of $\mb{K}$ involving near-degenerate states converges to zero in the limit of degeneracy.

Next, let us consider exactly degenerate states.
In the main text, we explained that if modes $m$ and $n$ degenerate, one needs to exclude $\gamma_{mn}$ from the set of variational parameters when studying the linear-response regime.
The reason is that the infinitesimal transformation parametrized by $\gamma_{mn}$ does not change the variational equilibrium density matrix $\oprho_0$.
One should exclude parameters that do not change the variational density matrix $\oprho(\mbx)$ at that given $\mbx$.
The parameter $\gamma_{mn}$ for degenerate $m$ and $n$ is such a parameter for the equilibrium, $\mbx=0$.
We note that such exclusion is necessary and justified only in the linear response regime.
When considering large deviations from $\mbx=0$, one must include $\gamma_{mn}$ in the set of variational parameters.

Even if the energy of degenerate states $m$ and $n$ are slightly perturbed due to numerical inaccuracies, no problem will occur.
As explained in the near-degenerate case, the corresponding equation of motion will be suppressed by a factor of $\sqrt{n_m-n_n}$, so that $\gamma_{mn}$ stays at is equilibrium value, $\gamma_{mn}=0$.

\section{Zero temperature case}
In the main text, we have focused only on the finite temperature case.
At zero temperature, one should apply TDVP directly to the Gaussian wavefunctions without purification.
The main difference with the finite temperature case is that the squeezing transformation parametrized by $\mb{\gamma}$ becomes a do-nothing operation at $T=0$.
This difference can be noticed by calculating the tangent vector by applying $\partial \opU / \partial \gamma$ [Eqs.~(\ref{eq:s_overlap_dU_cr}, \ref{eq:s_overlap_dU_ci})] to the stationary state wavefunction.
At $T > 0$, the purified stationary state wavefunction in the number basis has nonzero coefficients for states with nonzero phonon populations; hence, the tangent vectors do not vanish.
On the contrary, at $T=0$, the stationary state wavefunction is a vacuum state of the SCHA harmonic Hamiltonian.
Hence, the rightmost annihilation operators in Eqs.~(\ref{eq:s_overlap_dU_cr}, \ref{eq:s_overlap_dU_ci}) nullify the wavefunction and the corresponding tangent vectors become null vectors.
So, at zero temperature, only $\mb{\alpha}$ and $\mb{\beta}$ should be used as the variational parameters.

One can follow the same steps as in the finite temperature case to calculate the linearized time evolution generator and the position-position correlation function at zero temperature.
The final form of the phonon self-energy is identical to the finite-temperature result, \myeqref{eq:anh_Pi_chi}.
The only difference is that the second term in the definition of $\mb{\chi}$ [\myeqref{eq:anh_chi_def}] that originates from the variation of the $\mb{\gamma}$ parameter vanishes.
Still, the equations need not be modified because the second term of \myeqref{eq:anh_chi_def} is already zero at $T=0$ since $n_m = n_n = 0$.

\section{Single-mode anharmonic Hamiltonian} \label{sec:supp_single_mode}
In this section, we compute the excitation energy of the single-mode anharmonic Hamiltonian [\myeqref{eq:single_mode_ham}] using three different methods: perturbation theory, linearized time evolution, and projected Hamiltonian.

First, using standard second-order perturbation theory, the ground state and first-excited state energy are
\begin{equation}
    E_{\rm ground} = \frac{\omega_0}{2} - \frac{\lambda^2 a^2}{144\omega_0}
    + \mathcal{O}(\lambda^3)
\end{equation}
and
\begin{equation}
    E_{\rm 1st\ exc.} = \frac{3\omega_0}{2} - \frac{13 \lambda^2 a^2}{144\omega_0}
    + \mathcal{O}(\lambda^3).
\end{equation}
One can also show that the third-order perturbative correction to energy is zero because of the parity of the unperturbed wavefunctions.
Thus, the excitation energy is
\begin{equation} \label{eq:s_proj_w_pert}
    \omega_{\rm pert} = E_{\rm 1st\ exc.} - E_{\rm ground}
    = \omega_0 - \frac{\lambda^2 a^2}{12 \omega_0} + \mathcal{O}(\lambda^4).
\end{equation}

Next, let us use the linearized time evolution method.
The third- and fourth-order force constants of the Hamiltonian are
\begin{equation}
    \Phi^{(3)} = \lambda a, \quad \Phi^{(4)} = \lambda^2 b.
\end{equation}
Using the self-energy formula [\myeqref{eq:anh_Pi_chi}], we find
\begin{equation}
    \Pi(z) = - \frac{\omega_0 \lambda^2 a^2}{2(4\omega_0^2 - z^2)} \times \frac{1}{1 + \frac{\lambda^2 b\omega_0}{2(4\omega_0^2 - z^2)}}.
\end{equation}
The excitation energy $\omega_{\rm lin}$ is the position of the pole of the interacting Green function.
From the Dyson equation [\myeqref{eq:anh_Grr_Dyson}], one finds
\begin{equation}
    1 = \frac{\omega_0}{(\omega_{\rm lin})^2 - \omega_0^2} \Pi(\omega_{\rm lin}).
\end{equation}
In the perturbative limit of small $\lambda$, one finds
\begin{align} \label{eq:s_proj_w_lin}
    \omega_{\rm lin}
    \approx \omega_0 + \frac{1}{2} \Pi(\omega_0)
    = \omega_0
    - \frac{\lambda^2 a^2}{12\omega_0}
    + \mathcal{O}(\lambda^4).
\end{align}

Finally, we use the projected Hamiltonian method.
The tangent space of the Gaussian variational manifold at zero temperature is spanned by the 1- and 2-phonon states:
\begin{equation}
    \mathcal{T}_{\rm Gaussian} = {\rm span}\{ \ket{1}, \ket{2} \}.
\end{equation}
The Hamiltonian projected to this subspace is
\begin{equation}
    H_{\rm proj} =
    \begin{pmatrix}
    3\omega_0 / 2 & 0 \\
    0 & 5\omega_0 / 2
    \end{pmatrix}
    +
    \begin{pmatrix}
    0 & \lambda a/4 \\
    \lambda a/4 & \lambda^2 b / 8
    \end{pmatrix}.
\end{equation}
One can find the excitaiton energy by subtracting the variational ground state energy, $\omega_0/2$, from the lower eigenvalue of $H_{\rm proj}$:
\begin{align} \label{eq:s_proj_w_proj}
    \omega_{\rm proj}
    =& \frac{3 \omega_0}{2} + \frac{\lambda^2 b}{16}
    - \sqrt{ \left(\frac{\omega_0}{2} + \frac{\lambda^2 b}{16} \right)^2 + \left( \frac{\lambda a}{4} \right)^2 } - \frac{\omega_0}{2} \nnnl
    =& \omega_0 - \frac{\lambda^2 a^2}{16\omega_0} + \mathcal{O}(\lambda^4).
\end{align}

These results are summarized in Table~\ref{table:excitation} of the main text.
By comparing $\omega_{\rm lin}$ [\myeqref{eq:s_proj_w_lin}] and $\omega_{\rm proj}$ [\myeqref{eq:s_proj_w_proj}] to $\omega_{\rm pert}$ [\myeqref{eq:s_proj_w_pert}], we find that only the linearized time evolution method gives the correct leading order correction to the excitation energy.

\makeatletter\@input{xx.tex}\makeatother
\bibliography{main}